\documentclass[aps,amssymb,amsmath,pra,twocolumn,showpacs,superscriptaddress]{revtex4-1}

\pdfoutput=1
\usepackage{graphicx}% Include figure files
\usepackage{dcolumn}% Align table columns on decimal point
\usepackage{bm}% bold math
\usepackage{subfigure}
\usepackage{color}

\usepackage{amssymb,amsmath,amsfonts,latexsym,graphicx,verbatim}%,dsfont}

\usepackage[margin=0.5in]{geometry}

\usepackage[english]{babel}
\usepackage{times}
\usepackage{latexsym}
\usepackage{fancyhdr}
\usepackage{float}
\usepackage{afterpage}

\usepackage{multirow}
\usepackage{upgreek}

\definecolor{Ablue}{rgb}{0.96,0.24,0.00}

\definecolor{Abluetitle}{rgb}{0.,0.24,0.51}
\newcommand{\bluetitle}{\color{Abluetitle}}

\definecolor{orange}{rgb}{0.96,0.24,0.00}

\definecolor{darkred}{rgb}{0.55, 0.0, 0.0}

\definecolor{Gray}{gray}{0.85}
\definecolor{LightCyan}{rgb}{0.88,1,1}
\definecolor{darksalmon}{rgb}{0.91, 0.59, 0.48}
\definecolor{maroon}{cmyk}{0,0.87,0.68,0.32}

\definecolor{mustard}{rgb}{1.0, 0.86, 0.35}
\newcolumntype{a}{>{\columncolor{Gray}}c}
\newcolumntype{b}{>{\columncolor{white}}c}

\usepackage{array}
\newcolumntype{L}[1]{>{\raggedright\let\newline\\\arraybackslash\hspace{0pt}}m{#1}}
\newcolumntype{C}[1]{>{\centering\let\newline\\\arraybackslash\hspace{0pt}}m{#1}}
\newcolumntype{R}[1]{>{\raggedleft\let\newline\\\arraybackslash\hspace{0pt}}m{#1}}

\usepackage[colorlinks=true , citecolor=blue,urlcolor=blue]{hyperref}

\newcommand{\xd}{\delta}
\newcommand{\xe}{\epsilon}
\newcommand{\vxe}{\varepsilon}

\newcommand{\NV}{\R{NV}}
\newcommand{\CC}{\R{CC}}

\newcommand{\xg}{\gamma}
\newcommand{\xt}{\theta}

\newcommand{\xo}{\omega}

\newcommand{\xs}{\sigma}
\newcommand{\xz}{\zeta}
\newcommand{\pp}{\perp}
\newcommand{\app}{\approx}

\newcommand{\Bpol}{\textbf{B}_{\R{pol}}}
\newcommand{\Brel}{\textbf{B}_{\R{relax}}}
\newcommand{\Bp}{B_{\R{pol}}}
\newcommand{\Br}{B_{\R{relax}}}
\newcommand{\Cs}{{}^{13}\R{C}}
\newcommand{\Ns}{{}^{14}\R{N}}

\newcommand{\mV}[0]{\mathcal{V}}

\newcommand{\obs}{\R{obs}}

\newcommand{\xD}{\Delta}

\newcommand{\bz}[0]{\hat{\mathbf z}}

\newcommand{\fr}[2]{\frac{#1}{#2}}

\newcommand{\sq}[1]{\sqrt{#1}}

\newcommand{\mH}[0]{\mathcal{H}}

\newcommand{\rt}{\rightarrow}

\newcommand{\beq}{\begin{equation}}
\newcommand{\eeq}{\end{equation}}
                  
\newcommand{\benum}{\begin{enumerate}}
\newcommand{\eenum}{\end{enumerate}}
                    
\newcommand{\bit}{\begin{itemize}}
\newcommand{\eit}{\end{itemize}}

\newcommand{\bea}{\begin{eqnarray}}
\newcommand{\eea}{\end{eqnarray}}

\newcommand{\bev}{\begin{verbatim}}
\newcommand{\eev}{\end{verbatim}}

\newcommand{\non}{\nonumber}

\newcommand{\zt}{\times}

\newcommand{\qt}{\tau}

\newcommand{\lb}{\left(}
\newcommand{\rb}{\right)}
\newcommand{\lsb}{\left[}
\newcommand{\rsb}{\right]}
\newcommand{\lcb}{\left\{}
\newcommand{\rcb}{\right\}}
\newcommand{\labs}{\left|}
\newcommand{\rabs}{\right|}

\newcommand{\pll}{\parallel}

\newcommand{\T}[1]{\textbf{#1}}
\newcommand{\I}[1]{\textit{#1}}
\newcommand{\R}[1]{\textrm{#1}}
\newcommand{\zl}[1]{\label{eqn:#1}}
\newcommand{\zr}[1]{Eq. (\ref{eqn:#1})}
\newcommand{\zfl}[1]{\protect\label{fig:#1}}
\newcommand{\zfr}[1]{Fig. \ref{fig:#1}}

\newcommand{\zsl}[1]{\label{sec:#1}}
\newcommand{\zsr}[1]{Sec. \ref{sec:#1}}

\newcommand{\expec}[1]{\left\langle #1\right\rangle}

\newcommand{\ba}{\left\{ \begin{array}{lr}}
\newcommand{\ea}{\end{array}\right.}

\definecolor{darkred}{rgb}{0.55, 0.0, 0.0}
\newcommand{\Tr}[1]{\textrm{Tr}\left\{{#1}\right\}}

\newcommand{\blist}[1]{
 \begin{list}{#1}%$\ast\circ\bullet\Right
 \begin{align}
	 arrow
 \end{align}
 $\checkmark\star
  { \setlength{\itemsep}{3pt}
     \setlength{\parsep}{2pt}
     \setlength{\topsep}{3pt}
     \setlength{\partopsep}{0pt}
     \setlength{\leftmargin}{1em}
     \setlength{\labelwidth}{1em}
     \setlength{\labelsep}{0.5em} } }
\newcommand{\elist}{
  \end{list}  }

\DeclareMathSymbol{\vartheta}{\mathalpha}{letters}{"12}
\DeclareMathSymbol{\theta}{\mathalpha}{letters}{"23}
\DeclareMathSymbol{\phi}{\mathalpha}{letters}{"27}
\DeclareMathSymbol{\varphi}{\mathalpha}{letters}{"1E}

\newcommand{\bef}
{
\begin{figure}[htbp]
\centering
}

\newcommand{\eef}{\end{figure}}

\newcommand{\beginsupplement}{%
        \setcounter{table}{0}
        \renewcommand{\thetable}{S\arabic{table}}%
        \setcounter{figure}{0}
        \renewcommand{\thefigure}{S\arabic{figure}}%
     }

\newcommand{\affA}{ Department of Chemistry, and Materials Science Division Lawrence Berkeley National Laboratory University of California, Berkeley, California 94720, USA.}
\newcommand{\affB}{Department of Physics and CUNY-Graduate Center, CUNY-City College of New York, New York, NY 10031, USA.}

\newcommand{\affD}{Department of Chemical and Biomolecular Engineering, and Materials Science Division Lawrence Berkeley National Laboratory University of California, Berkeley, California 94720, USA.}
\newcommand{\affE}{Fakult\"{a}t Physik, Technische Universit\"{a}t Dortmund, D-44221 Dortmund, Germany.}

\newcommand{\affH}{Department of Physics and Astronomy, Dartmouth College, Hanover, New Hampshire 03755, USA.}

\begin{document}
\title{\bluetitle{Hyperpolarized relaxometry based nuclear $T_1$ noise spectroscopy in hybrid diamond quantum registers}}

\author{A. Ajoy}\email{ashokaj@berkeley.edu}\affiliation{\affA}
 \author{B. Safvati}\affiliation{\affA}
  \author{R. Nazaryan}\affiliation{\affA}
   \author{J. T. Oon}\affiliation{\affA}
   \author{B. Han}\affiliation{\affA}
   \author{P. Raghavan}\affiliation{\affA}
   \author{R. Nirodi}\affiliation{\affA}
   \author{A. Aguilar}\affiliation{\affA}
\author{K. Liu}\affiliation{\affA}
  \author{X. Cai}\affiliation{\affA}
   \author{X. Lv}\affiliation{\affA}
   \author{E. Druga}\affiliation{\affA}
   \author{C. Ramanathan}\affiliation{\affH}
   \author{J. A. Reimer}\affiliation{\affD}
   \author{C. A. Meriles}\affiliation{\affB}
   \author{D. Suter}\affiliation{\affE}
   \author{A. Pines}\affiliation{\affA}

%\author{Authors}\affiliation{\affA}

\begin{abstract}
 Understanding the origins of spin lifetimes in hybrid quantum systems is a matter of current importance in several areas of quantum information and sensing. Methods that spectrally map spin relaxation processes provide insight into their origin and can motivate methods to mitigate them. In this paper, using a combination of hyperpolarization and precision field cycling over a wide range (1mT-7T), we map frequency dependent relaxation in a prototypical hybrid system of $\Cs$ nuclear spins in diamond coupled to Nitrogen Vacancy centers. Nuclear hyperpolarization through the optically pumped NV electrons allows signal time savings for the measurements exceeding million-fold over conventional methods. We observe that $\Cs$ lifetimes show a dramatic field dependence, growing rapidly with field up to $\sim$100mT and saturating thereafter. Through a systematic study with increasing substitutional electron (P1 center) concentration as well as $\Cs$ enrichment levels, we identify the operational relaxation channels for the nuclei in different field regimes. In particular, we demonstrate the dominant role played by the $\Cs$ nuclei coupling to the interacting P1 electronic spin bath. These results pave the way for quantum control techniques for dissipation engineering to boost spin lifetimes in diamond, with applications ranging from engineered quantum memories to hyperpolarized $\Cs$ imaging.
\end{abstract}

\maketitle

\begin{figure}[t]
  \centering
  \includegraphics[width=0.49\textwidth]{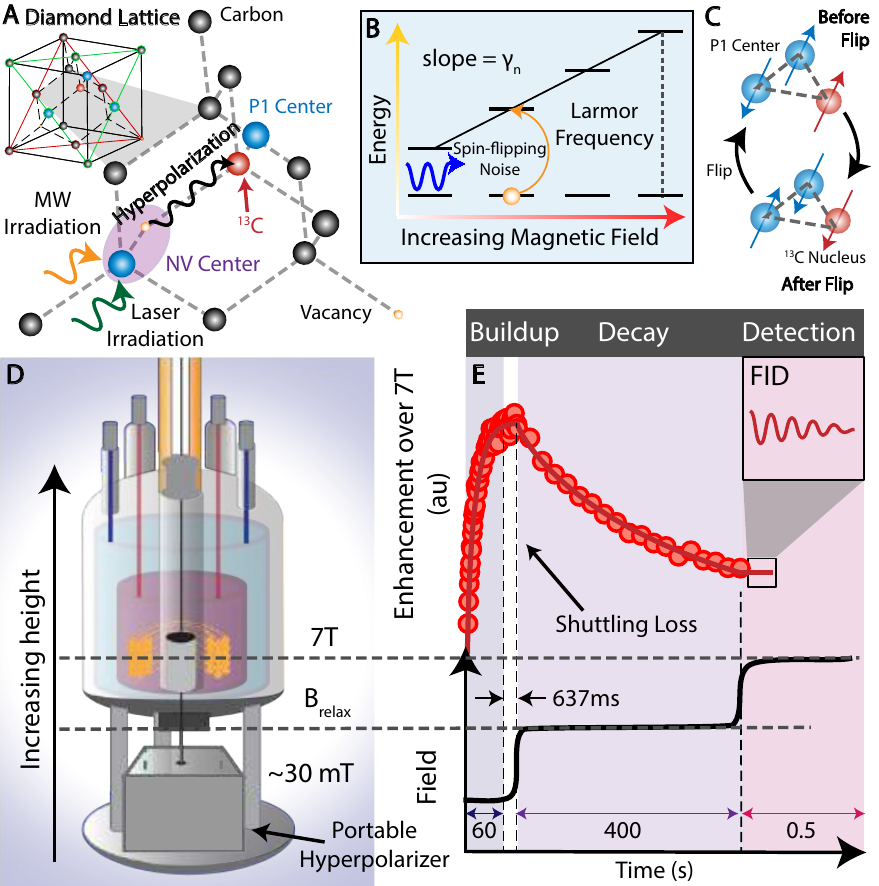}
  \caption{\T{Principle.} (A) \I{System} consisting of $\Cs$ nuclear spins in diamond hyperpolarized via NV centers allowing their direct measurement by bulk NMR. Lattice also contains electronic spin bath of P1 centers. (B) Changing magnetic field allows probing of spin flipping noise that is resonant with the carbon Larmor frequency. (C) Dominant $T_1$ relaxation mechanism via three-body flip-flops with pairs of P1 center electrons. (D) \I{Experimental platform.} Portable hyperpolarizer is installed in a rapid field cycling device capable of sweeping between 10mT-7T in the fringe field of a NMR magnet. (E) \I{Time sequence.}  Lower panel shows the schematic steps of laser driven optical $\Cs$ hyperpolarization for $\sim$60s at $\Bp\app$30mT, rapid shuttling ($<$1s) to the field of interest $\Br$, relaxation and subsequent high field detection at 7T.  Upper panel displays typical data for 200$\mu$m microdiamond powder, where $\Br=$ 2T. $\Cs$ NMR signal amplitude (points) is quantified by its enhancement over the 7T Boltzmann signal. Signal growth and decays are fitted to stretched exponentials (solid lines).}
\zfl{schematic}
\end{figure}

\T{\I{Introduction:}} -- The power of quantum technologies, especially those for information processing and metrology, relies critically on the ability to preserve the fragile quantum states that are harnessed in these applications~\cite{Preskill98}. Indeed noise serves as an encumbrance to practical implementations, causing both decoherence as well as dissipation of the quantum states~\cite{Zurek03,zollerBook}. Precise \I{spectral} characterization of the noise opens the door to strategies by which it can be effectively suppressed~\cite{Alvarez11,Suter2016} -- case in point being the emergence of dynamical decoupling techniques that preserve quantum coherence by periodic driving~\cite{viola00}. In these cases, quantum control sets up a filter that decouples components of noise except those resonant with the exact filter period~\cite{Cywinski08}, allowing spectral decomposition of the \I{dephasing} noise afflicting the system. Experimentally implemented in ion traps~\cite{Biercuk09}, superconducting qubits~\cite{Bylander11} and solid-state NMR~\cite{Ajoy11}, this has spurred development of Floquet engineering to enhance $T_2$ decoherence times by over an order of magnitude in these physical quantum device manifestations~\cite{Ryan10, Gustavsson12, Bar-Gill12}.

Methods that analogously spectrally fingerprint $T_1$ \I{relaxation} processes, on the other hand, are more challenging to implement experimentally. If possible however, they could reveal the origins of relaxation channels, and foster means to suppress them. Applications to real-world quantum platforms are pressing: relaxation in Josephson junctions and ion trap qubits, for instance, occur due to often incompletely understood interactions with surface paramagnetic spins~\cite{Labaziewicz08}. Relaxation studies are also important in the context of hybrid quantum systems, such as those built out of coupled electronic and nuclear spins. In the case of diamond Nitrogen Vacancy (NV) center electronic qubits coupled to $\Cs$ nuclei~\cite{Jelezko06}, for instance, a detailed understanding of nuclear relaxation can have important implications for quantum sensing~\cite{Degen17}: engineered NV-$\Cs$ clusters form building blocks of quantum networks~\cite{Taminiau12}, are the basis for spin gyroscopes~\cite{Ajoy12g}, and are harnessed as quantum memories in high-resolution nano-MRI probes~\cite{Rosskopf17}. Nuclear $T_1$ lifetimes are not dominated by phonon interactions, but instead are set by couplings with the intrinsic electronic spin baths themselves -- a complex dynamics that is often difficult to probe experimentally. Indeed only a small proportion of $\Cs$ spins can be addressed or readout via the NV centers, as also the direct inductive readout of these spins suffer from extremely weak signals. Moreover, as opposed to $T_2$ noise spectroscopy carried out in the rotating frame~\cite{Bar-Gill12}, probing of $T_1$ processes have to be performed in the laboratory frame. This necessitates the ability to probe relaxation behavior while subjecting samples to widely varying magnetic field strengths.

In this paper, we develop a method of \I{``hyperpolarized relaxometry''} that overcomes these instrumentational and technical challenges. We measure  $T_1$ relaxation rates of $\Cs$ spins in diamond samples relevant for quantum sensing with a high density of NV centers. Our $T_1$ noise spectroscopy proceeds with high resolution and over four decades of noise spectral frequency, revealing the physical origins of the relaxation processes. While experiments are demonstrated on diamond, it acts here as a prototypical solid state electron-nuclear hybrid quantum system, and the results are indicative of relaxation processes operational in other systems, including Si:P~\cite{Morton08}, wide bandgap materials such as SiC~\cite{christle15,klimov15}, and diamond-based quantum simulator platforms constructed out of 2D materials such as graphene and hBN~\cite{Cai13,Lovchinsky17,Ajoy19}. These results are also pertinent for producing and maintaining polarization in hyperpolarized solids, for applications employing hyperpolarized nanoparticles of Si or diamond as MRI tracers~\cite{Cassidy13,Wu16}, and in the relayed  optical DNP of liquids mediated through nanodiamonds~\cite{Ajoy17}, since in these applications $T_1$ relaxation bounds the achievable polarization levels. 

Key to our technique is the hyperpolarization of $\Cs$ nuclei at room temperature, allowing the rapid and direct measurement of nuclear spin populations via bulk NMR~\cite{Ajoy17}.  Dynamic nuclear polarization (DNP) is carried out by optical pumping and polarizing the NV electrons (close to 100\%) and subsequently transferring polarization to $\Cs$ nuclei (\zfr{schematic}A). This routinely leads to nuclear polarization levels $\gtrsim$0.5\%. In a high-field (7T) NMR detection spectrometer, for instance, the signals are enhanced by factors exceeding $\vxe\sim$300-800 times the Boltzmann value~\cite{Ajoy17}, boosting measurement times by 10$^5$-10$^6$, and resulting in high single shot detection SNRs. This permits $T_1$ spectroscopy experiments that would have otherwise been intractable.  Hyperpolarization is equally efficiently generated in single crystals as well as randomly oriented diamond powders, and both at natural abundance as well as enriched $\Cs$ concentrations. The hyperpolarized samples are interfaced to a home built field cycler instrument~\cite{Ajoyinstrument18}  (see \zfr{schematic}D and video in ~\cite{shuttlervideo}) that is capable of rapid and high-precision changes in magnetic field over a wide 1mT-7T range (extendable in principle from 1nT-7T), opening a unique way to peer into the origins of nuclear spin relaxation.

\begin{figure}[t]
  \centering
  \includegraphics[width=0.5\textwidth]{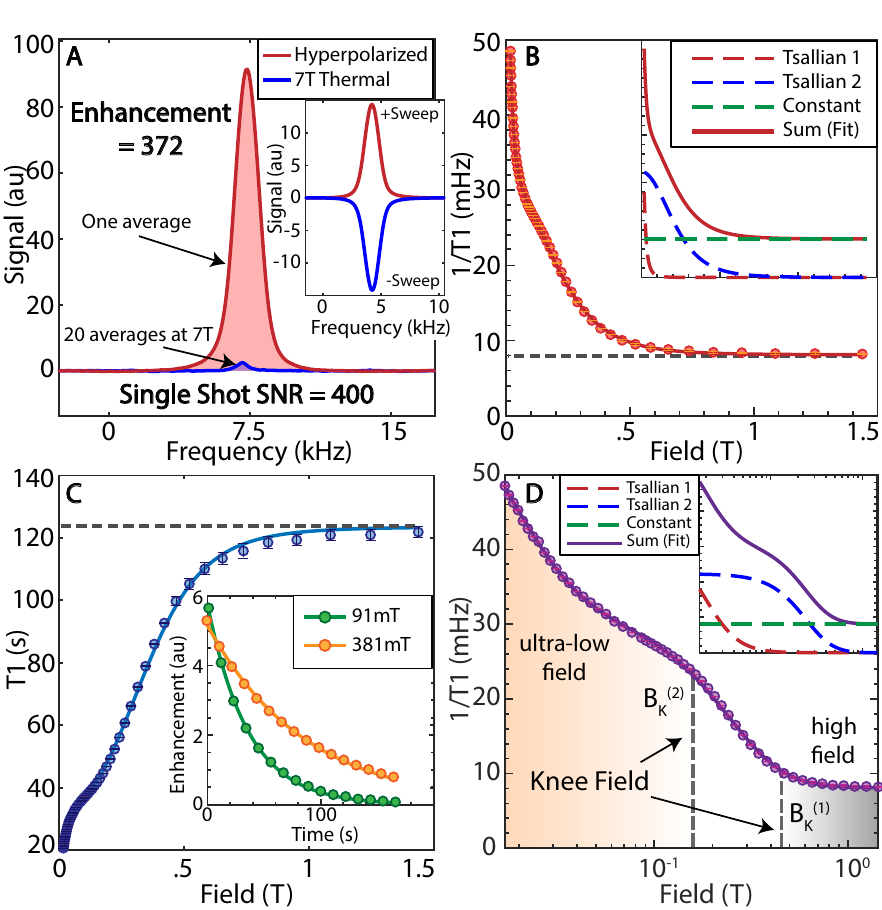}
  \caption{\textbf{Hyperpolarized relaxometry} applied to a 10\% $\Cs$ enriched single crystal. (A) \I{Signal gains} due to hyperpolarization under optimal conditions at $\Bp\app$36mT. Red line shows a single-shot hyperpolarized signal (SNR$\app$400) after 60s of optical pumping. Blue line is the 7T thermal signal after 20 averages, allowing us to quantify signal enhancement from DNP $\app$372 over 7T, a time saving by $\app10^6$ for equal SNR. \I{Inset:} Exemplary signals at $\Bp\app$36mT under low-to-high (high-to-low) frequency sweeps leading to positive (negative) $\Cs$ hyperpolarization. (B) \I{Relaxation rate} $R_1=1/T_1$ obtained from relaxometry over a wide field range 20mT-1.5T. We observe a rapid growth in  relaxation rate below a knee field of 0.5T, and saturation at higher fields. \I{Inset:} Data can be fit to two Tsallian functions, which we ascribe to be originating from inter-$\Cs$ couplings and interactions to the P1 spin bath. (C) \I{Spin lifetimes} as a function of field,  showing significant boost in nuclear $T_1$ beyond the knee field, approaching a lifetime $\app$2.1min. \I{Inset:}  Typical relaxation data at two representative fields showing monoexponential character. (D) \I{Logarithmic scale} data visualization, displaying a more equanimous sampling of experimental points, and the knee fields inflection points $B_{K}^{(1,2)}$.  \I{Inset:} Decomposition into the constituent Tsallians. Error bars in all panels are obtained from monoexponential fits. }
  \zfl{ten}
\end{figure}

\T{\I{$\Cs$ Hyperpolarized relaxometry:}} -- \zfr{schematic}D-E schematically describe the experiment. Hyperpolarization in the $\Cs$ nuclei is affected by optical pumping at low fields, typically $\Bp\sim$40mT, followed by rapid transfer to the intermediate field $\Br$ where the spins are allowed to thermalize (see \zfr{schematic}C), and subsequent bulk inductive measurement at 7T. Experimentally varying $\Br$ allows one to probe field dependent lifetimes $T_1(\Br)$, and through them noise sources perpendicular to $\Brel$ and {resonant} with the nuclear Larmor frequency $\xg_n\Br$ (\zfr{schematic}B). Here $\xg_n=10.7$MHz/T is the $\Cs$ gyromagnetic ratio. This allows the spectral decomposition of noise processes that spawn $T_1$ relaxation. For instance pairs of substitutional nitrogen impurities (P1 centers) undergoing flip-flops (\zfr{schematic}C) can apply on the $\Cs$ nuclei a stochastic spin-flipping field that constitutes a relaxation process.

Optical excitation for hyperpolarization involves 520nm irradiation at low power ($\sim$80mW/mm$^2$) applied continuously for $\sim$40s. Microwave (MW) sweeps, simultaneously applied across the NV center ESR spectrum, transfer this polarization to the $\Cs$ spins (see \zfr{ten}A)~\cite{Ajoy17, Ajoy18}.   DNP occurs in a manner that is completely independent of crystallite orientation. All parts of the underlying NV ESR spectrum produce hyperpolarization, with intensity proportional to the underlying electron density of states. The polarization sign depends solely on the direction of  MW sweeps through the NV ESR spectrum (see \zfr{ten}A inset). Physically, hyperpolarization arises from partly adiabatic traversals of a pair of Landau-Zener (LZ) crossings in the rotating frame that are excited by the swept MWs. For a more detailed exposition of the DNP mechanism, we point the reader to Ref.~\cite{Zangara18}.

Low field hyperpolarization is hence excited independent of the fields $\Br$ under which relaxation dynamics is to be studied. There is significant acceleration in acquisition time since optical DNP obviates the need to thermalize spins at high fields where $T_1$ times can be long (for some samples $>$30min). Gains averaging time are $\app\vxe^2 \fr{T_{1}(\R{7T})}{T_1(\Bp)}$, which in our experimental conditions exceeds five orders of magnitude. In \zfr{ten}A for instance on a 10\% enriched single crystal, we obtain large DNP enhancements $\vxe=$380, and high single shot SNR $\app$400. It also reflects the inherently high DNP efficiency: every NV center has surrounding it $\sim$10$^5$ nuclear spins, which we polarize to a bulk value (averaged over all $\Cs$ nuclei) of 0.37\% employing just 3000 MW sweeps, indicating a transfer efficiency of $\app$12.3\% per sweep per fully polarized nuclear spin. Harnessing this large signal gain allows us to perform relaxometry at a density of field points that are about two orders of magnitude greater than previous efforts~\cite{Reynhardt03a,lee11}. Such high-resolution spectral mapping (for instance 55 field points in \zfr{ten}) can transparently reveal the underlying processes driving nuclear relaxation. Indeed, in future work, use of small flip-angle pulses might allow one to obtain the entire relaxation curve with a single round of DNP, and thus the \I{ultrafast relaxometry} of the nuclei. 

Our experiments are also aided by technological attributes of the DNP mechanism. DNP is carried out under low fields and laser and MW powers, and allows construction of a compact hyperpolarizer device that can accessorize a field cycling instrument~\cite{Ajoy18pol} (see ~\cite{hypercubevideo} for video of hyperpolarizer operation). The wide range (1mT-7T) field cycler is constructed over the 7T detection magnet, and affects rapid magnetic field changes by physically transporting the sample in the axial fringe field environment of the magnet~\cite{Ajoyinstrument18}. This is accomplished by a fast (2m/s) conveyor belt actuator stage (Parker HMRB08) that \I{shuttles} the sample via a carbon fiber rod (see video in Ref.~\cite{shuttlervideo}). The entire sample (field) trajectory can be programmed, allowing implementation of the polarization, relaxation and detection periods as in \zfr{schematic}C. Transfer times at the maximum travel range were measured to be 648$\pm$4ms~\cite{SOM}, short in comparison with the $T_{1n}$ lifetimes we probe. High positional resolution (50$\mu$m) allows access to field steps at high precision (~\cite{SOM} shows full field-position map). The field is primarily in the $\bz$ direction (parallel to the detection magnet), since sample transport occurs centrally, and the diameter of the shuttling rod (8mm) is small in comparison with the magnet bore (54mm).

\T{\I{Results:}} -- \zfr{ten} shows representative results of $T_1$ noise spectroscopy on $\Cs$ nuclei in diamond, considering here 
a 10\% enriched single crystal.  The intriguing data can be visualized in several complementary ways. First, considering relaxation rate $R_1=1/T_1$ (\zfr{ten}B), the high-resolution data allows us to clearly discern three regimes: a steep narrow $R_1$ increase at ultralow fields ($<$10mT), a broader component at moderate fields (10mT-500mT), and an approximately constant relaxation rate independent of field beyond 0.5T and extending upto 7T (data beyond 2T not shown). Each point in \zfr{ten}B reports the monoexponential decay constant obtained from the \I{full} decay curve at every field value (for example shown in \zfr{ten}C). Error bars at each field value are estimated from monoexponential fits of the polarization decays. The resulting errors are under a few percent. The solid line in \zfr{ten}B indicates a numerical fit and remarkably closely follows the experimental data. Here we employ a sum of two Tsallian functions~\cite{Tsallis95,Howarth03} that capture the decay rates at low and moderate fields, and a constant offset at high field (see \zfr{ten}B insets). 

A second viewpoint of the data, presented in \zfr{ten}C, is of the $T_1$ relaxation times and highlights its highly nonlinear field dependence. There is a \I{step}-like behavior in $T_1(\Br)$, and an inflection point (\I{knee} field) $\app$100mT beyond which the $T_1$'s saturate. We quantify the knee field value,  $B^{(1)}_{K}$, as the $\Br$ at which the relaxation rate is twice the saturation $R_1$ that we observe at high field. This somewhat counterintuitive dependence has significant technological implications. \I{(i)} Long $\Cs$ lifetimes can be fashioned even at relatively modest fields at room temperature. This adds value in the context of $\Cs$ hyperpolarized nanodiamonds as potential MRI tracers~\cite{Rej15}, since it provides enough time for the circulation and binding of surface functionalized particles to illuminate disease conditions. \I{(ii)} The step-behavior in \zfr{ten}C also would prove beneficial for $\Cs$ hyperpolarization storage and transport. Exceedingly long lifetimes can be obtained by simply translating polarized diamond particles to modest $\sim$100mT fields -- low enough to be produced by simple permanent magnets~\cite{Ajoy18pol}.  

Finally, while the visualizations in \zfr{ten}B,C cast light on the low and high field behaviors respectively,  the most natural representation of the wide-field data is on a logarithmic scale (\zfr{ten}D). The high-density data now unravels the rich relaxation behavior at play in the different field regimes.  We discern an additional second inflection point $B^{(2)}_{K}$ at lower magnetic fields below which there is a sudden increase in the relaxation rates. The inset in \zfr{ten}D shows the decomposition into constituent Tsallian fits with a narrow and broad widths. %As we shall demonstrate, these point to  due to $\Cs$ interactions with lattice electronic spin baths at moderate fields, and inter-nuclear dipolar couplings at ultalow fields. 

Microscopic origins of this relaxation behavior can be understood by first considering the diamond lattice to consist of three disjoint spin \I{reservoirs} -- electron reservoirs of NV centers, P1 centers, and the $\Cs$ nuclear spin reservoir.  P1 centers arise predominantly during NV center production on account of finite conversion efficiency in the diamond lattice. Indeed the P1 centers are typically at 10-100 times higher concentration than NV centers; with typical lattice concentrations of NVs, P1s and $\Cs$ nuclei respectively $P_{\R{NV}}\sim$1ppm, $P_e\sim$10-100ppm, and $P_C\sim 10^4\eta\:$ppm, where $\eta$ is the $\Cs$ lattice enrichment level. At any non-zero field of interest, $\Br$, the electron and nuclear reservoirs are centered at widely disparate frequencies and do not overlap. We can separate the relaxation processes in different field regimes to be driven respectively by -- \I{(i)} couplings of $\Cs$ nuclei to pairs (or generally the reservoir) of P1 centers. This leads to the $B^{(1)}_{K}$ feature at moderate fields in \zfr{ten}C; \I{(ii)} $\Cs$ spins interacting with individual P1 or NV centers undergoing lattice driven relaxation ($T_{1e}$ processes); \I{(iii)} inter-nuclear couplings within the $\Cs$ reservoir that convert Zeeman order to dipolar order. Both of the latter processes contribute to the low field $B^{(2)}_{K}$ features in \zfr{ten}C; and finally, \I{(iv)} a high-field process $>$1T that shows a slowly varying (approximately constant) field profile. We ascribe this to arise directly or indirectly (via electrons) from two-phonon Raman processes. Since these individual mechanisms are independent, the overall relaxation rate is obtained through a sum, $\fr{1}{T_1} = \sum_{(J)}\fr{1}{T_1^{(J)}}$ (shown in the inset of \zfr{ten}D).

\begin{figure*}[t]
  \centering
  \includegraphics[width=0.95\textwidth]{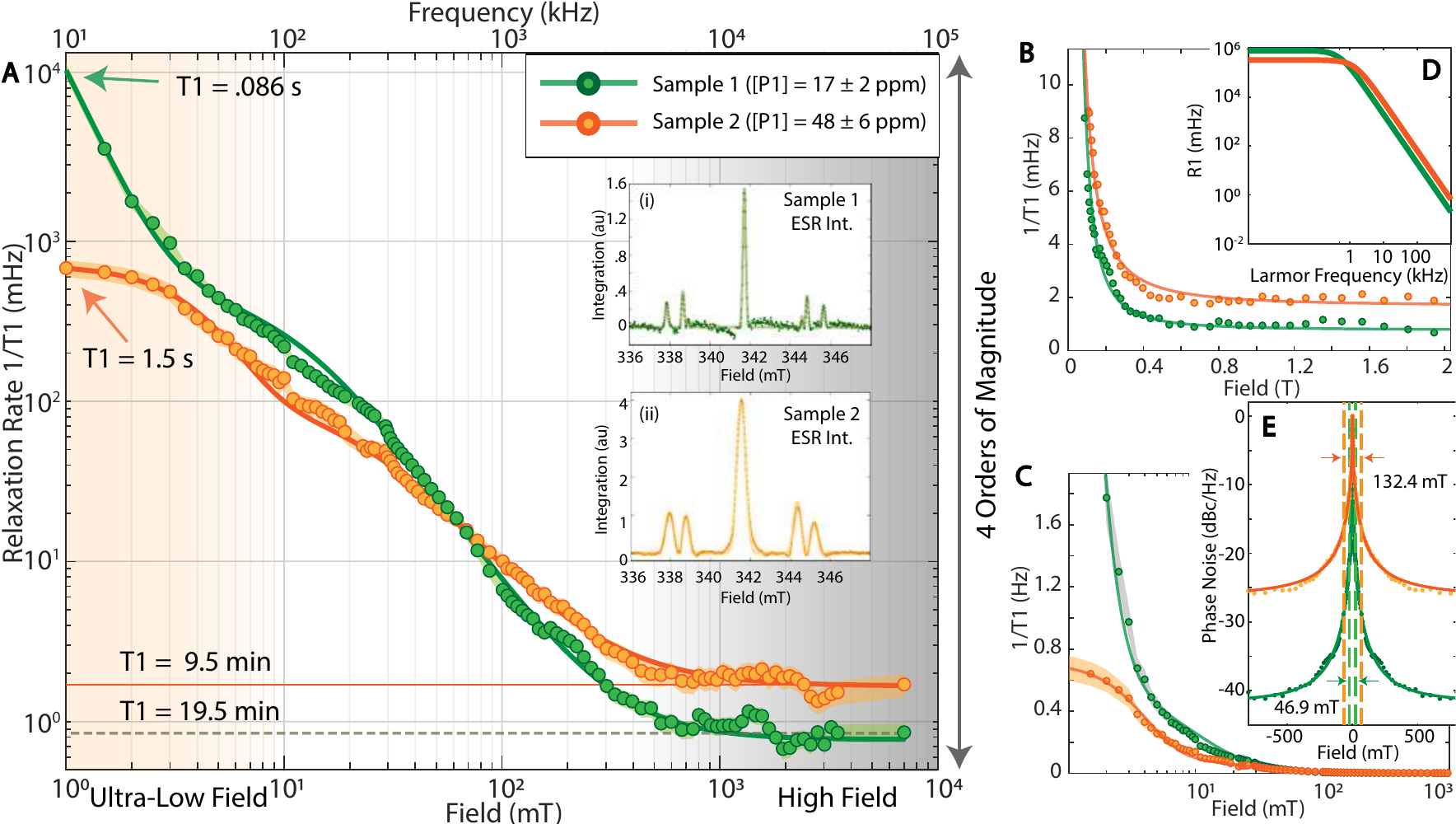}
  \caption{\T{Hyperpolarized relaxometry at natural abundance} $\Cs$ concentration over four decades of field 1mT-7T (lower axis), probing spin-flipping spectral density from 1kHz-75MHz (upper axis). (A) \I{Relaxation rate} on a logarithmic scale, showing steep field dependence that spans four orders of magnitude in $T_1$, falling to sub-second lifetimes at ultra-low fields below $B_{K}^{(2)}$, and saturating to lifetimes greater than 10min. beyond $B_{K}^{(1)}$. Orange and green data correspond to CVD samples with different concentration of P1 centers~\cite{Scott16} (legend). Solid lines are fits to a combination of two Tsallian functions. Shaded regions represent error bounds originating from our accelerated data collection strategy (see Supplemental Information~\cite{SOM}). \I{Insets:} X-band ESR spectra. (B) \I{High field behavior} shows saturating knee field $B_{K}^{(1)}$ occurs at higher field for sample with larger P1 concentration. (C) \I{Low field behavior}, where intriguingly sample with more P1 centers has a lower relaxation rate. (D) \I{Calculated relaxation rate} $R_1(\xo_L)$ arising from the coupling of the $\Cs$ spins with the interacting P1 reservoir for the case of 17ppm (green) and 48ppm (orange) electron concentrations, showing qualitative agreement with the experimental data. (E) \I{Comparing effective phase noise} $S_p(\xo)$ for the two samples on a semi-log scale. For clarity, data is mirrored on the X-axis and phase noise normalized against relaxation rates at $\xo_0$=1mT.  Solid lines are fits to Tsallian functions. Dashed vertical lines indicate the theoretical widths obtained from the the respective estimates of $\expec{d_{ee}}$, 46.7mT and 131.89mT, matching very closely with experiments. }
  \zfl{one_percent}
\end{figure*}

\T{\I{Effect of electronic spin bath:}} -- Let us first experimentally consider the relaxation process stemming from $\Cs$ spins coupling to the interacting P1 reservoir. In \zfr{one_percent} we consider single crystal samples of natural $\Cs$ abundance grown under similar conditions but with different nitrogen concentrations. Their P1 electron concentrations are $P_e=$17ppm and $P_e=$48ppm, measured from X-band ESR~\cite{Drake16} (shown in \zfr{one_percent}D). To obtain data with high density of field-points, hyperpolarized relaxometry measurements are taken by an {accelerated} strategy (outlined in ~\cite{SOM}) over a ultra-wide field range from 1mT-7T, with DNP being excited at $\Bp$=36mT. For relaxometry at fields below $\Bp$, we employ rapid current switching of Helmholtz coils within the hyperpolarizer device. Both the range of fields, as well as the density of field-points being probed are significantly higher than previous studies~\cite{Reynhardt03a,Reynhardt97}. This aids in quantitatively unraveling the underlying physics of the relaxation processes. We note that probing relaxation behavior below $\sim$1mT in our experiments is currently limited by the finite sample shuttling time, which becomes of the order of the $T_1$'s being probed. 

Experimental results in \zfr{one_percent} reveal a remarkably sharp $R_1$ dependence, best displayed in \zfr{one_percent}A on a logarithmic scale, showing variation in relaxation rate over four orders of magnitude. The data fits two Tsallian functions (solid line), and reveals the $B^{(1)}_{K}$ inflection point (closely resembling \zfr{ten}B) beyond which the lifetimes saturate. The second knee field $B^{(2)}_{K}$ at ultralow fields can also be discerned, although determining its exact position is difficult without relaxation data approaching truly zero-field. Comparing the two samples (\zfr{one_percent}A), we observe a clear correlation in the $B^{(1)}_{K}$ knee field values shifting to higher fields at higher electron concentration $P_e$. The high field relaxation rates, highlighted in \zfr{one_percent}B, increase with $P_e$. Interestingly at low fields (see \zfr{one_percent}C), the sample with lower $P_e$ has an enhanced relaxation rate, yielding an apparent ``cross-over'' in the relaxation data between the two samples at $\app$50mT. While we have focused here on single crystals, we observe quantitatively identical relaxation behavior also for microdiamond powders down to 5$\mu$m sizes (see \zfr{powder}).  This is because the random orientations of the crystallites play no significant role in the P1-driven nuclear relaxation process. We do expect, however, that for nanodiamond particles $<$100nm, surface electronic spins will cause an additional relaxation channel.

Let us now develop a simple model to quantify this P1-dominated relaxation process. Given the low relative density of the NV centers and consequently weak NV-NV couplings, to a good approximation they play no role except to inject polarization into the $\Cs$ nuclei. Consider the Hamiltonian of the \I{system}, assumed for simplicity to be a single $\Cs$ spin, and the \I{environment} - the interacting bath of P1 centers surrounding it,
$
\mH = \mH_S + \mH_E + \mH_{SE} + \mH_{EE}
$
where, the first two terms capture the Zeeman parts, the third term is the coupling between reservoirs, and the last term captures the inter-electron dipolar couplings within the P1 bath. Specifically,
\bea
\mH &=& \xo_LI_{z} + \xo_eS_z + \sum_jA_{zx}^j S_{zj}I_{x} \non\\
&+& \sum_{j<k} d^{ee}_{jk}\lsb S_{zj}S_{zk} + \fr{1}{2}\lb S_j^+S_k^- + S_j^-S_k^+\rb\rsb.
\eea
where $I$ (and $S$) refer to spin-$\fr{1}{2}$ Pauli operators on the nuclei (electrons) respectively, and $A_{zx}^j$ the pseudo-secular hyperfine interaction that can drive nuclear spin-flips on the $\Cs$ nuclei. For simplicity, we neglect here the effect of the P1 hyperfine couplings to host $\Ns$ nuclei. In principle, they just split the electronic reservoirs seen by the $\Cs$ nuclei into three manifolds separated by the large hyperfine coupling $A_{\pll}^{\R{P1}}\app$114MHz. In the rotating frame at $\mH_{E}$, and going into an interaction picture with respect to $\mH_{EE}$, the Hamiltonian becomes,
$
\mH_I =  \xo_LI_{z} + \expec{A_{zx}} I_x\sum_j\lb e^{-i\mH_{EE}t} \hat{S}_{z}e^{i\mH_{EE}t}\rb 
=  \xo_LI_{z} +\expec{A_{zx}}\hat{S}_z(t)I_{x}
$
with,
$\expec{A_{zx}}=\sq{\expec{A_{zx}^2}} = \lsb\sum_j \lb A_{zx}^j\rb ^2\rsb^{1/2}$ and the operator $\hat{S}_z = \fr{1}{\expec{A_{zx}}}\sum_j A_{zx}^j S_{zj}$. Here $\expec{A_{zx}}$ is the total effective P1-$\Cs$ hyperfine interaction, and the norm $\|\mH_{EE}\|$ is set by the average dipolar interaction between electronic spins in the bath, henceforth $\expec{d_{ee}}$. We now make a semi-classical approximation, promoting $\hat{S}_z$ to $s_z(t)$, a variable that represents a classical stochastic process seen by the $\Cs$ nuclear spins~\cite{Abragam61, Ajoy11},
\beq
\mH_I = \xo_LI_{z} + \expec{A_{zx}}s_z(t)I_{x}\:.
\zl{fluc}
\eeq
In summary, a spin flipping term $I_x$ is tethered to a stochastic variable $s(t)$ and this serves as ``noise'' on the $\Cs$ spins, flipping them at random instances and resulting in nuclear relaxation upon a time (or ensemble) average. Interestingly, this noise process arises due to electronic flip-flops in the \I{remote} P1 reservoir that is widely separated in frequency from $\Cs$ spins.  In a simplistic picture, shown in \zfr{schematic}C, relaxation originates from \I{pairs} of P1 centers in the same $\Ns$ nuclear manifold (energy-mismatched by $\xd$) undergoing spin flip-flop processes, and flipping a $\Cs$ nuclear spin (when $\xo_L\app\xd$) in order to make up the energy difference. In reality, the overall relaxation is constituted out of several such processes over the entire P1 electronic spectrum.

Let us now assume the stochastic process $s_z(t)$ is Gaussian with zero mean and an autocorrelation function $g(\qt)=\exp(-\qt/\qt_c)$ with  correlation time $\qt_c = 1/\expec{d_{ee}}$. The spectral density function $S(\xo)=\fr{1}{\sq{2\pi}}\int_{\infty}^{\infty}g(\qt)e^{-i\xo\qt}d\qt$ that quantifies the power of the spin flipping noise components at various frequencies is then a Lorentzian, $S(\xo)=2\qt_c/(1+\xo^2\qt_c^2)$. Going further now into an interaction picture with respect to $\xo_LI_z$, 
$
\mH_{I}^{(I)}=\expec{A_{zx}}s_z(t)\lb e^{-i\xo_LI_zt'}I_xe^{i\xo_LI_zt'}\rb.
$
The survival probability of the spin is,
$
p(t)=\fr{1}{2}\Tr{I_z e^{i\mH_{I}^{(I)}t}I_ze^{-i\mH_{I}^{(I)}t}}\sim e^{-\chi(t)}
$
where in an average Hamiltonian approximation, retaining effectively time-independent terms, the effective relaxation rate $\chi(t)\app R_1t$ can be obtained by \I{sampling} of the spectral density resonant with the nuclear Larmor frequency $\xo_L$ at each field point. This is the basis behind noise spectroscopy of the underlying $T_1$ process~\cite{Kimmich04}. We recover then the familiar Bloembergen-Purcell-Pound (BPP) result~\cite{bloembergen48,Redfield57}, where the relaxation rate,
\beq
\fr{1}{T_1^{(1)}} = R_1^{(1)}(\xo_L) = \expec{A_{zx}^2}S(\xo_L) = \expec{A_{zx}^2}\fr{\expec{d_{ee}}}{\xo_L^2 + \expec{d_{ee}}^2}
\zl{electron_relax}
\eeq

The inter-spin couplings can be estimated from the typical inter-spin distance $\expec{r_e}=\lb 3/4\pi \ln 2\rb^{1/3}N_e^{-1/3}$, where  $N_e=(4\zt 10^{-6}P_e)/a^3$[m$^{-3}$] is the electronic concentration in inverse volume units and $a$=0.35nm the lattice spacing in diamond~\cite{Reynhardt03a}. The couplings are now related to the second moment of the electronic spectra~\cite{Abragam61}
$
M_{2e} = \fr{9}{20} (g\mu_B)^2\fr{1}{\expec{r_e}^6},
$
where $g\app 2$ is the electron g-factor, and $\mu_B=9.27\zt10^{-21}$erg/G the Bohr magneton in cgs units. This gives $\expec{d_{ee}}\app \xg_e\sq{\fr{8}{\pi}}\sq{M_{2e}}\:$[Hz]$\app$10.5$P_e$[mG], which scales approximately linearly with electron concentration $P_e$~\cite{Reynhardt03a}. For the two samples with $P_e$=17ppm and 48ppm we obtain spectral widths $\expec{d_{ee}}$=0.5kHz and 1.42kHz respectively, corresponding to  field-profile widths of 46.9mT and 132.4mT respectively. These would correspond to inflection points $B_{K}^{(1)}= \fr{\expec{d_{ee}}}{2\xg_n}$ in the relaxometry data at fields 23.5mT and 66.2mT respectively. These values, represented by the dashed lines in \zfr{one_percent}A, are in remarkable quantitative agreement with the experimental data. Moreover, we expect that these turning points (scaling $\propto P_e$) are \I{independent} of $\Cs$ enrichment $\eta$, in agreement with the data in \zfr{ten} (see also \zfr{enriched}). 

From lattice considerations (see~\cite{SOM}), we can also estimate the value of the effective hyperfine coupling  $\expec{A_{zx}}$ in \zr{electron_relax}, which we expect to grow slowly with $P_e$. We make the assumption that there is barrier of $r_0\app$2.15nm around every P1 center in which the $\Cs$ spins are ``unobservable'' because their hyperfine shifts exceed the measured $\Cs$ linewidth $\xD f_{\R{det}}\app$2kHz. Our estimate can be accomplished by sitting on a P1 spin, and evaluating $\expec{A_{zx}} = \lsb\expec{A_{zx}^2}\rsb^{1/2}$, where the second moment~\cite{Abragam61},
$
\expec{A_{zx}^2} = \fr{1}{N}\lsb \fr{\mu_0}{4\pi}\xg_e\xg_n\hbar\rsb^2\sum_{j}\fr{(3\sin\xt_j\cos\xt_j)^2}{r_j^6}
$ with $N$ being the relative number of $\Cs$ spins per P1 spin, and $\xt_j$ the angle between the P1-$\Cs$ axis and the magnetic field, and index $j$ runs over the region between neighboring P1 spins.  This gives,
\beq
\expec{A_{zx}^2} \app \lb \fr{\mu_0}{4\pi}\xg_e\xg_n\hbar\rb^2 \fr{6}{5} \fr{1}{\expec{r_e}^3}\lb \fr{1}{r_0^3} - \fr{1}{\expec{r_e}^3}\rb
\zl{hyp_eqn}
\eeq
 For the two samples, we have $\expec{r_e}=$ 4.8nm and 3.39nm respectively, giving rise to the effective P1-$\Cs$ hyperfine interaction $\expec{A_{zx}^2}\app$ 0.39[(kHz)$^2$] and $\expec{A_{zx}^2}\app$0.45[(kHz)$^2$] respectively. These values are also consistent with direct numerical estimates from simulated diamond lattices (see ~\cite{SOM}). The simple model stemming from \zr{fluc} and \zr{electron_relax} therefore predicts that the effective hyperfine coupling $\expec{A_{zx}}$ increases slowly with the electron concentration $P_e$,  with the electron spectral density width $\expec{d_{ee}}\propto P_e$.

% In order to validate the conclusions from this simple model, we perform an alternative numerical estimation of $\expec{A_{zx}^2}=\lsb \fr{1}{N}\sum_{j\in \xD f_{\R{det}}}\expec{A^2_{zx,j}}\rsb$ within the detection barrier directly from the diamond lattice (see Supplemental Information~\cite{SOM}). We obtain $\expec{A_{zx}^2}=$2[(kHz)$^2$] and 2.26[(kHz)$^2$] for the $P_e$=17ppm and 48ppm samples respectively, in quantiative agreement with the values predicted from \zr{hyp_eqn}.

Finally, from \zr{hyp_eqn} we can estimate the zero-field rate stemming from this relaxation process,  $R_1(0) = \fr{\expec{A_{zx}^2}}{\expec{d_{ee}}}\app$ 777[s$^{-1}$] and 317.5[s$^{-1}$] respectively. \zfr{one_percent}D calculates the resulting relaxation rates from this process $R_1(\xo_L)$ in a logarithmic plot. It shows good semi-quantitative agreement with the data in \zfr{one_percent}A and captures the experimental observation that the rates of the two samples ``cross over'' at a particular field. It is instructive to represent the data in terms of effective ``phase noise'' (see \zfr{one_percent}E), denoted logarithmically as, $S_p(\xo_L) = 10\log\lb\fr{R_1(\xo_0)}{R_1(\xo_L)}\rb$[dBc/Hz], where $\xo_0\rt 0$ represents the relaxation rates approaching zero field. \zfr{one_percent}E shows this for the two samples, employing $\xo_0=$1mT and with the estimated field-linewidths displayed by the dashed lines. This makes evident that the high field spin-flipping noise seen by the $\Cs$ nuclei is about 15dB lower in the 17ppm sample.

While \zr{electron_relax} is the dominant relaxation mechanism operational at moderate fields, let us now turn our attention to the the behavior at ultralow fields in \zfr{one_percent}. \zr{fluc} provides the framework to consider the effect of \I{single} P1 and NV electrons to the relaxation of $\Cs$ nuclei. In this case the stochastic process $s_z(t)$ arises not on account of inter-electron couplings, but due to individual $T_{1e}$ processes operational on the electrons, due to for instance coupling to lattice phonons. The width of the spectral density is then given by $T_{1e}$, 
\beq
\fr{1}{T_1^{(2)}} = R_1^{(2)}(\xo_L) = \expec{A_{zx}^2}\fr{T_{1e}}{1+\xo_L^2T_{1e}^2}
\zl{electron_relax2}
\eeq
While $T_{1e}$ is also field-dependent, and dominated by two-phonon Raman processes at moderate-to-high field, typical values of $T_{1e}\sim$1ms~\cite{Jarmola12}, give rise to Lorentzian relaxometry widths of $\app$1kHz, corresponding to field turning points of $B_{K}^{(2)}\app\fr{1}{2\xg_nT_{1e}}=$0.1mT. 

\begin{figure}[t]
  \centering
  \includegraphics[width=0.5\textwidth]{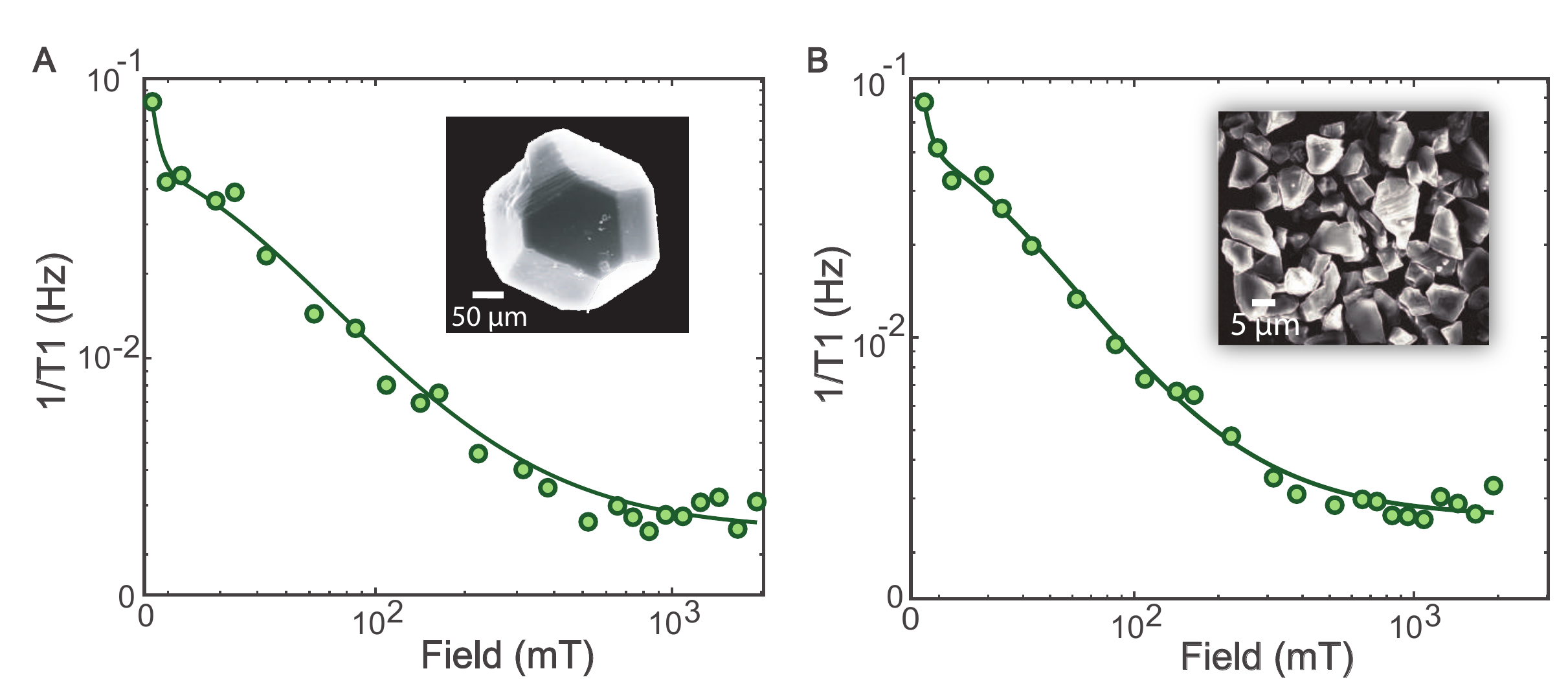}
  \caption{\T{$\Cs$ nuclear relaxation in microdiamond powder}. Relaxation field maps for the randomly oriented natural abundance $\Cs$ microdiamond powders of size (A) 200$\mu$m and (B) 5$\mu$m with accompanying SEM images (insets). Data is obtained by measuring the full relaxation curve at every field point, and is quantitatively similar to the single crystal results in \zfr{one_percent}.}
  \zfl{powder}
\end{figure}

\begin{figure}[t]
  \centering
  \includegraphics[width=0.46\textwidth]{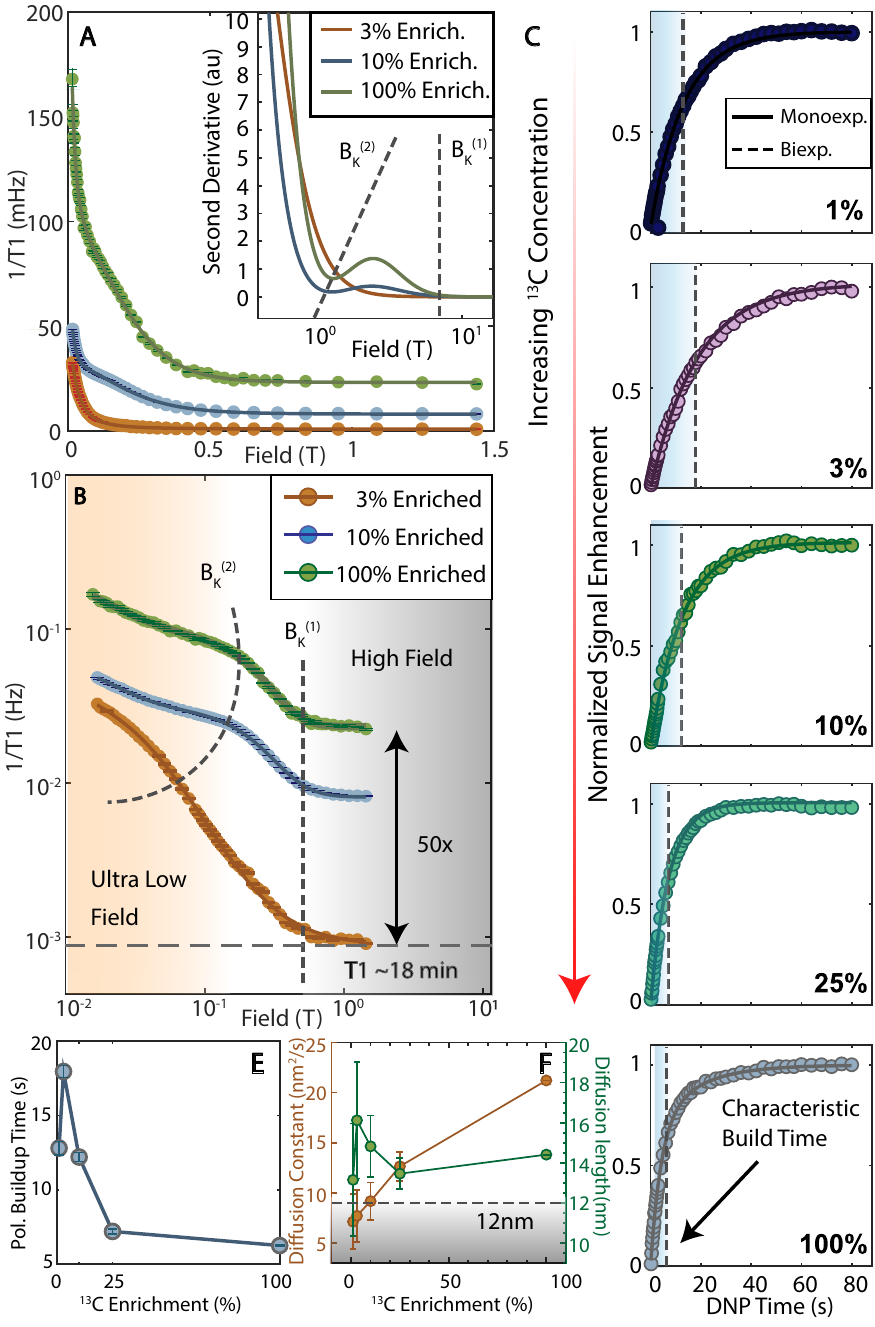}
  \caption{\T{Variation with $\Cs$ enrichment}. Experiments are performed on single crystal samples placed so that all the NV center orientations are identical at 54.7$^{\circ}$ to $\Bp$=36mT. (A) \I{Relaxation rates} on linear and (B) logarithmic field scale, making evident an increase in relaxation rate with increasing $\Cs$ enrichment at low and high fields. Solid lines are Tsallian fit. Error bars are obtained from the relaxation data at various fields. Characteristic knee field $B_{K}^{(1)}$ (dasher vertical line) at moderate fields is independent of enrichment, evident in the inset. Knee field at ultra-low fields $B_{K}^{(2)}$ qualitatively is indicated by the dashed line that serves as a guide to the eye. \I{Inset:} Second derivative of the fitted lines, showing the knee fields at the zero-crossings. (C) \I{DNP polarization buildup curves} also reflect differences in the nuclear spin lifetimes, displaying saturation at much shorter times upon increasing enrichment. DNP in all curves are performed at 36mT sweeping the entire $m_s=+1$ manifold. (D) Polarization buildup times extracted from the data showing that  faster nuclear spin relaxation limits the final obtained hyperpolarization enhancements in highly enriched samples. (F) \I{Spin diffusion constant and diffusion length} for $\Cs$ nuclei numerically estimated from the data as a function of lattice enrichment. Dashed line indicates the mean inter-electron distance $\expec{r_{\NV}}\app$12nm between NV centers at 1ppm concentration, indicating that spin diffusion can homogeneously spread polarization in the lattice almost independent of $\Cs$ enrichment. }
  \zfl{enriched}
\end{figure}

\T{\I{Effect of $\Cs$ enrichment:}} --  To systematically probe this low-field behavior as well as consider the effect of couplings within the $\Cs$ reservoir, we consider in \zfr{enriched} diamond crystals with varying $\Cs$ enrichment $\eta$ and approximately identical NV and P1 concentrations. With increasing enrichment, a third relaxation mechanism becomes operational, wherein at low fields it becomes possible to dissipate Zeeman energy into the dipolar bath. The field dependence of this process is expected to be more Gaussian, centered at zero field and have a width $\sim \expec{d_{\CC}}$ the mean inter-spin dipolar coupling between $\Cs$ nuclei. We can estimate (see~\cite{SOM}) these couplings from the second moment, 
$
\expec{d_{\CC}} = \fr{1}{N}\sum_j\lsb\sum_k \lb \fr{\mu_0}{4\pi}\hbar\xg_n^2(3\cos^2\xt_{jk}-1)\rb^2/{r_{jk}^6}\rsb^{1/2}
$
where in a lattice of size $\ell$, $N=N_C\ell^3$ refers to the number of $\Cs$ spins, and the spin density $N_C=0.92\eta$ spins/nm$^3$. Here $\xt_{jk}=\cos^{-1}\lb \fr{\T{r}_{jk}\cdot\Brel}{r_{jk}\Br}\rb$ is the angle between the inter-nuclear vector and the direction of the magnetic field. In the numerical simulations (outlined in ~\cite{SOM}), we evaluate the case consistent with experiments wherein the single crystal samples placed flat, i.e. with $\Brel\pll$ [001] crystal axis. As a result, for $\Cs$ spins on adjacent (nearest-neighbor) lattice sites, $\xt_{jk}=$54.7$^{\circ}$ is the magic angle and $d_{jk}^{\CC}=0$. 

We find $\expec{d_{\CC}}\app$850Hz for natural abundance samples and a scaling $\expec{d_{\CC}}\propto\eta^{1/2}$ with increasing enrichment. This is in good agreement with the experimentally determined linewidths (see ~\cite{SOM}). We thus expect a turning point at low fields, $B_{K}^{(2)}\sim \fr{\expec{d_{\CC}}}{2\xg_n}$, for instance $\app$39$\mu$T for natural abundance samples, but scaling to $\app$0.46mT in case of the 100\% enriched sample. In real experiments, it is difficult to distinguish between this process and that arising directly from single electrons in \zr{electron_relax2}, and hence we assign the same label to this field turning point.

Performing hyperpolarized relaxometry (see \zfr{enriched}) we observe that increasing enrichment leads to a fall in nuclear $T_1$s, evident both at low (\zfr{enriched}A) and high (\zfr{enriched}B) fields. $R_1$ rates for the highly enriched samples (10\% and 100\%) are obtained by taking the full relaxation decay curves at every field point, while for the low enriched sample (3\%) enrichment, we use an accelerated data collection strategy (see~\cite{SOM}) on account of the inherently long $T_1$ lifetimes.  On a logarithmic scale (\zfr{enriched}B), we observe the knee field $B^{(1)}_{K}$ is virtually identical across all the samples, indicating it is a feature independent of $\Cs$ enrichment, originating from interactions with the electronic spin bath. This is in good agreement with the model in \zr{electron_relax}. A useful means to evaluate the inflection points from the zeros of the second derivative of the Tsallian fits, as indicated in the inset of \zr{electron_relax}A. Moreover, the lower inflection field $B^{(2)}_{K}$ scales to higher fields with increasing enrichment $\eta$, pointing to its origin from \I{internuclear} dipolar effects. At the low fields, we also notice that the samples with lower enrichment have higher relaxation rates, and with steeper field-profile slopes (\zfr{enriched}B). This is once again consistent with the model that the spectral density height and width being probed scales  with $\expec{d_{\CC}}$.  %We speculate that this maybe due to an Anderson localization effect~\cite{Bar-Gill12} in high $\eta$ samples, the $\Cs$ nuclei being disordered in frequency due to stronger hyperfine couplings to the electronic spin bath, and the energy mismatch suppressing flip-flops between them.

Changes in the nuclear lifetimes are also reflected directly in the DNP polarization buildup curves, shown in \zfr{enriched}C. We perform here hyperpolarization of all the samples under the same conditions, sweeping the entire $m_s$=+1 manifold at $\Bp$=36mT, sweeping over the full NV ESR spectrum.  We notice that polarization buildup is predominantly mono-exponential (dashed lines in \zfr{enriched}C), except for at natural abundance $\Cs$, where a biexponential growth (solid line) is indicative of nuclear spin diffusion.  Data demonstrates that highly enriched samples have progressively smaller polarization buildup times (see \zfr{enriched}E) on account of limited nuclear lifetimes at $\Bp$.

Moreover, the experimental data allows us to quantify the "`homogenization"' of polarization in the lattice. We assign a spin diffusion coefficient $D= \fr{\expec{r_n}^2}{30T_{2n}}$ (see \zfr{enriched}F) where the $T_{2n}$ are evaluated here by only taking the dipolar contribution to the linewidth, $T_{2n}\app 1/\expec{d_{\CC}}$~\cite{Hayashi08}. Given a total time bounded by $T_1$, we can calculate the rms overall diffusion length~\cite{Zhang98} as $\xs=\sq{2DT_1}$ that is displayed as the blue points in \zfr{enriched}F. Also for reference is plotted the mean NV-NV distance $\app$12nm at 1ppm concentration (dashed region in \zfr{enriched}F), indicating that to a good approximation the optically pumped polarization reaches to all parts of the diamond lattice between the NV centers.

We comment finally that determining the origins of $\Cs$ relaxation in enriched samples can have several technological applications. Enrichment provides an immediate means to realize quantum registers and sensing modalities constructed out of hybrid NV-$\Cs$ spin clusters,  and as such ascertaining nuclear relaxation profiles is of practical importance for such applications. Low $\eta$ ($\leq$ 3\%) naturally engender NV-$\Cs$ pairs that can form quantum registers~\cite{Dutt07,Neumann10,Reiserer16}. The nuclear spin can serve as an ancillary quantum memory that, when employed in magnetometry applications, can provide significant boosts in sensing resolution~\cite{Laraoui13,Rosskopf17}.  With increasing $\Cs$ concentrations $\eta\gtrsim$10\% a single NV center can be coupled to several $\Cs$ nuclei forming natural nodes for a quantum information processor, and where the nuclear spins can be actuated directly by hyperfine couplings to the NV electron~\cite{Khaneja07,Borneman12}. Approaching full enrichment levels ($\eta=$100\%), internuclear couplings become significant, permitting hybridized nuclear spin states and decoherence protected subspaces~\cite{kalb17} for information storage. In bulk quantum sensing too, for instance applied to diamond based gyroscopes~\cite{Ajoy12g,Maclaurin12}, the high density of $\Cs$ sensor spins ($\sim 10^{22}$/cm$^{3}$), as much as $>10^5$ times the number of NV centers, can be harnessed to increase sensitivity. % Ascertaining factors that affect the $\Cs$ lifetimes in such samples is therefore of significant practical importance. 

\begin{figure}[t]
  \centering
  \includegraphics[width=0.4\textwidth]{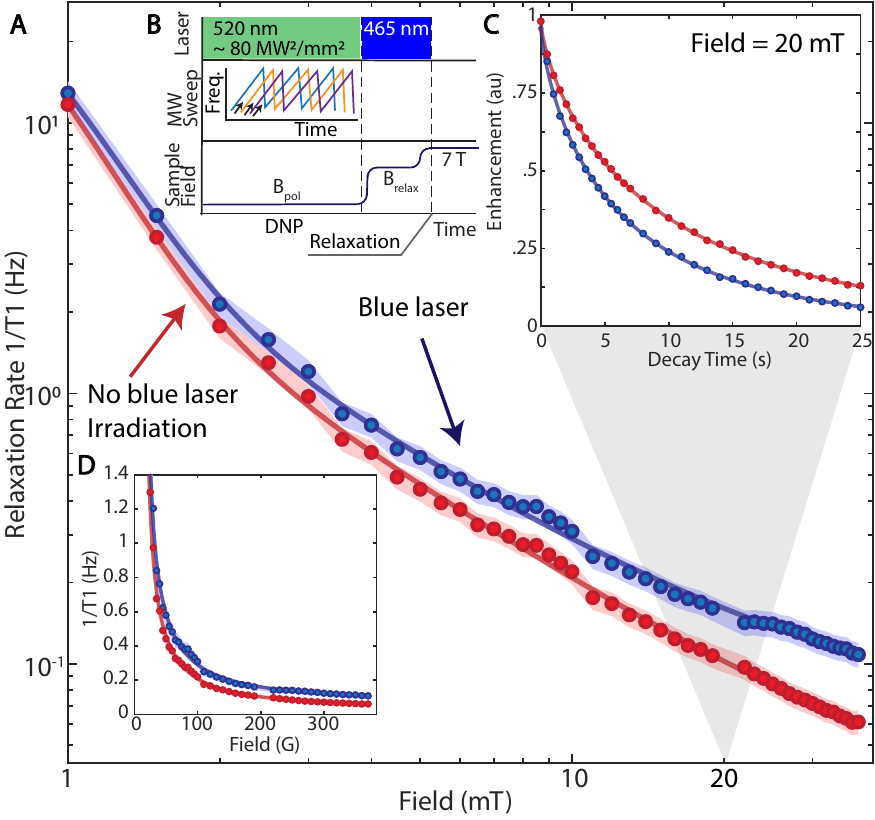}
  \caption{\T{Dynamic optical engineering of electron spin density.} Panel denotes $\Cs$ relaxation rate on a logarithmic scale for 17ppm P1 concentration sample employed in \zfr{one_percent} with relaxation dark (red points) and under low power (80mW/mm$^2$) blue (465nm) irradiation (blue points). Shaded regions represent error bounds (see~\cite{SOM}). (B) \I{Schematic} time sequence of the experiment. (C) Exemplary decay curves obtained at 20mT. (D) Relaxation rates on linear field scale.  We observe that the blue radiation leads to a decrease in $\Cs$ nuclear lifetimes, which we hypothesize arises from fluctuations introduced in the electronic spin bath upon recapture after P1 center ionization. This illustrates that the electronic spin spectral density can be optically manipulated, and potentially ultimately also narrowed under sufficiently high-power ionization irradiation.}
  \zfl{blue}
\end{figure}

\T{\I{Discussion:}} -- Experimental results in \zfr{one_percent} and \zfr{enriched} substantiate the $\Cs$ relaxation pathways operational at different field regimes, and potentially highlight the particularly important role played by the electronic reservoir towards setting the spin lifetimes. Our work therefore opens the door to a number of intriguing future directions.  First, it suggests the prospect of increasing nuclear lifetimes by raising the NV center conversion efficiency~\cite{Farfurnik17}.  More generally, it points to the efficacy of materials science approaches towards reducing paramagnetic impurities in the lattice. Finally, it opens the possibility of employing coherent quantum control for dissipation engineering, to manipulate the spectral density profile seen by the nuclei and consequently lengthen their $T_1$. Applying a ``\I{pulse sequence to increase $T_1$}'' has been a longstanding goal in magnetic resonance~\cite{Carravetta04,Pileio10}, but is typically intractable because of inability to coherently control broad-spectrum phonon interactions. Instead here since the nuclear $T_1$ stems from electronic $T_{2e}$ processes, these can be ``echoed out''; In particular, the application of electron decoupling (such as WAHUHA~\cite{Waugh68} or Lee-Goldburg~\cite{Lee65} decoupling) on the P1 spin bath would suppress the inter-electron flip-flops, narrow the noise spectral density, and consequently shift the knee field $B^{(1)}_{K}$ to lower fields. Such $T_1$ gains just by spin driving at room temperature and without the need for cryogenic cooling, and consequent boosts in the hyperpolarization enhancements -- scaling by the decoupling factor -- will have far-reaching implications for the optical DNP of liquids under ambient conditions. 

Given the multi-frequency microwave control driving each of the $\Ns$ manifolds would entail~\cite{Bauch18}, an attractive alternate \I{all-optical} means is via the optical ionization of P1 centers, for instance by  irradiation at blue ($\lesssim$ 495nm) wavelengths where the P1 electrons ionize strongly~\cite{Aslam13}. Sufficiently rapid electronic ionization, faster than their flip-flop rate, would once again narrow the spectral density and increase nuclear $T_1$. \zfr{blue} shows preliminary experiments in this direction, where we study the change in the relaxation rate under 465nm blue irradiation. Due to technical limitations (sample heating) we limit ourselves to the low power $\sim$80mW/mm$^2$ regime. We observe a comparative \I{decrease} in nuclear $T_1$ with respect to decay in the dark. Note that, in contrast, we do not observe significant change in the lifetimes under 520nm excitation. Under weak blue excitation the P1 centers are not ionized fast enough, and we hypothesize that upon electron recapture, the P1 centers can affect the $\Cs$ nuclei over a longer distance in a lattice. The blue irradiation thus causes a ``\I{stirring}'' of the electronic spin bath and an increase in the nuclear relaxation rate. While the experiments in \zfr{blue} unambiguously affirm that interactions with the electronic bath set the low field nuclear $T_1$, the exact interplay between optical ionization and recapture rates required for $T_1$ suppression is a subject we will consider in future work.

\T{\I{Conclusions:}} -- Employing hyperpolarized relaxometry, we have mapped the $\Cs$ nuclear spin lifetimes in a prototypical diamond quantum system over a wide field range, in natural abundance and enriched $\Cs$ samples, and for both single crystals as well as powders. We observe a dramatic and intriguing field dependence, where spin lifetimes fall rapidly below a knee field of $\sim$100mT. The results indicate that the spin lifetimes predominantly arise from nuclear flip processes mediated by the P1 center electronic spin bath, and immediately opens the compelling possibility of boosting nuclear lifetimes by quantum control or optically induced electronic ionization. This has significant implications in quantum sensing, in building longer lived quantum memories, and in practically enhancing the $\Cs$ hyperpolarization efficiency in diamond, with applications to hyperpolarized imaging of surface functionalized nanodiamonds and for the DNP of liquids brought in contact with high surface area diamond particles. 

\T{\I{Acknowledgments:}} -- We gratefully acknowledge discussions with A. Redfield, D. Sakellariou and J.P. King, and technical contributions from  M. Gierth, T. McNelly, and T. Virtanen. C.A.M. acknowledges support from NSF through NSF-1401632 and NSF-1619896, from Research Corporation for Science Advancement through a FRED award, and research infrastructure from NSF Centers of Research Excellence in Science and Technology Center for Interface Design and Engineered Assembly of Low-Dimensional Systems (NSF-HRD-1547830).

\T{\I{Materials}} -- $\Cs$ enriched diamonds used to conduct experiments in \zfr{ten} and \zfr{enriched} were grown through chemical vapor deposition using a $\Cs$  enrichment mixture of methane and nitrogen (660ppm, Applied Diamond Inc) as precursor followed by $\Cs$  enrichments of 10\%, 25\%, 50\%, and 100\% to produce the respective percent-enriched diamonds~\cite{Parker17}.
To produce a NV-concentration of 1-10ppm, the enriched samples were irradiated with 1MeV electrons at a fluence of 10$^{18}$ cm$^{-2}$ then annealed for 2 hours at 800C.  The natural abundance samples used in \zfr{one_percent} were grown under synthetic high pressure, high temperature conditions (Element 6, Sumitomo)~\cite{Scott16} then annealed for 1 hour at 850$^{\circ}$C.  The NV and P1 concentration were measured to be 1.4$\pm$0.02ppm and 17$\pm$2ppm for the first sample and 6.9$\pm$0.8ppm and 48$\pm$6ppm for the second sample, respectively. The microdiamond powders in \zfr{powder}, produced by HPHT techniques, were acquired respectively from Element6 and Columbus Nanoworks.

%\bibliography{C:/paper-drafts/Biblio}
\bibliography{Biblio}
\bibliographystyle{apsrev4-1}

\pagebreak
\clearpage
\onecolumngrid
%\begin{widetext}
\begin{center}
\textbf{\large{\textit{Supplementary Information} \\\smallskip
\bluetitle{Hyperpolarized relaxometry based nuclear $T_1$ noise spectroscopy in hybrid diamond quantum registers}}}\\
\hfill \break
\smallskip
A. Ajoy,$^{1,\textcolor{red}{\ast}}$  B. Safvati,$^{1}$ R. Nazaryan,$^{1}$  J. T. Oon,$^{1}$ B. Han,$^{1}$ P. Raghavan,$^{1}$ R. Nirodi,$^{1}$ A. Aguilar,$^{1}$ K. Liu,$^{1}$\\ X. Cai,$^{1}$ X. Lv,$^{1}$ E. Druga,$^{1}$ C. Ramanathan,$^{2}$ J. A. Reimer,$^{3}$ C. A. Meriles,$^{4}$ D. Suter,$^{5}$ and A. Pines$^{1}$\\
\smallskip
\emph{${}^{1}$ {\small Department of Chemistry, University of California Berkeley, and Materials Science Division Lawrence Berkeley National Laboratory, Berkeley, California 94720, USA.}}
\emph{${}^{2}$ {\small Department of Physics and Astronomy, Dartmouth College, Hanover, New Hampshire 03755, USA.}}
\emph{${}^{3}$ {\small Department of Chemical and Biomolecular Engineering, and Materials Science Division Lawrence Berkeley National Laboratory University of California, Berkeley, California 94720, USA.}}
\emph{${}^{4}$ {\small Department of Physics, and CUNY-Graduate Center, CUNY-City College of New York, New York, NY 10031, USA.}}
\emph{${}^{5}$ {\small Fakultat Physik, Technische Universitat Dortmund, D-44221 Dortmund, Germany.}}
\end{center}

% \begin{minipage}[t]{0.625\textwidth}
% \begin{center}
% \end{center}
% \end{minipage}
% \end{center}

%\end{widetext}

\twocolumngrid

\beginsupplement
{ \hypersetup{linkcolor=darkred}
\tableofcontents}

\section{EPR Measurements in \zfr{one_percent}}
EPR spectra of the two samples in \zfr{one_percent} were examined with a microwave power 6mW, averaging over 50 sweeps, with modulation amplitudes of 0.1mT and 0.01mT and at sweep fields of 3350G - 3500G and 3300G - 3600G for the two samples respectively. Concentrations of P1 centers were estimated by using a CuSO$_4$ reference outlined in Ref.~\cite{Scott16}. 

In order to determine the linewidths of the EPR spectra, a script was written to determine the data range at which Tsallis fits should be applied by first finding the indices where the spectral maxima
and minima occured. Midpoints were then determined between the maximum and minimum indices and the first derivative of the Tsallis function was fit to the ranges between the calculated midpoints. Because the baseline was not perfectly zeroed, jumps in the fit values occurred between each range. Applying fits to each individual peak rather than applying one Tsallis function to multiple peaks produced a better baseline correction since the offsets differed between ranges. Each peak was corrected by subtracting the median y-value over the fit range and then making manual corrections if necessary. Once the corrections were completed, the first integrals over each individual range were obtained using trapezoidal integration. The resulting integral arrays were then concatenated and a second integral was obtained. The resulting
first integral allowed us to find the line widths of each P1 peak (FWHMs), and the second integral resembled a step function from which the relative step heights of each P1 peak could be found. To
account for the hyperfine splittings of the P1 spectra an average over all peaks linewidths was taken and weighted by the height of each peak. The ratio of the averaged linewidths between the two
samples in \zfr{one_percent} was found to be 2.97, consistent with the ratio of the P1 concentration of the two samples up to the accuracy of the concentration estimates.
\begin{figure}[t]
  \centering
  \includegraphics[width=0.5\textwidth]{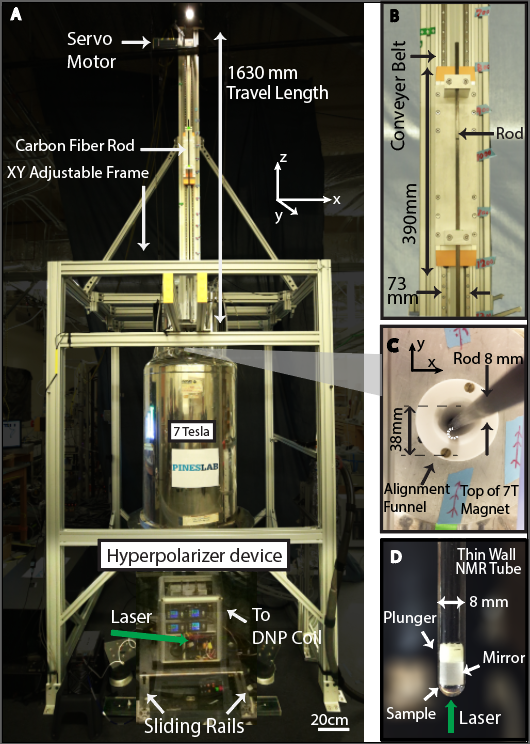}
  \caption{\textbf{Field cycling device} interfaced with portable diamond hyperpolarizer. (A)  Mechanical shuttler is connected to a 7T magnet and interfaced to a portable hyperpolarizer. Sliding rails attached to the bottom of the device allow for adjustment of hyperpolarizer box and centering of sample above coil. (B) The carbon-fiber shuttling rod is moved along a conveyor belt through use of a twin-carriage actuator. (C) The 8mm shuttling rod is centered in the 38mm magnet bore, with a Teflon guide for self-alignment. (D) Diamond sample is held within an 8mm wide NMR tube, and fitted with a plunger and mirror to prevent excess movement of sample and bolster efficacy of optical pumping.}
  \zfl{field_cycler}
\end{figure}

\begin{figure}[t]
  \centering
  \includegraphics[width=0.3\textwidth]{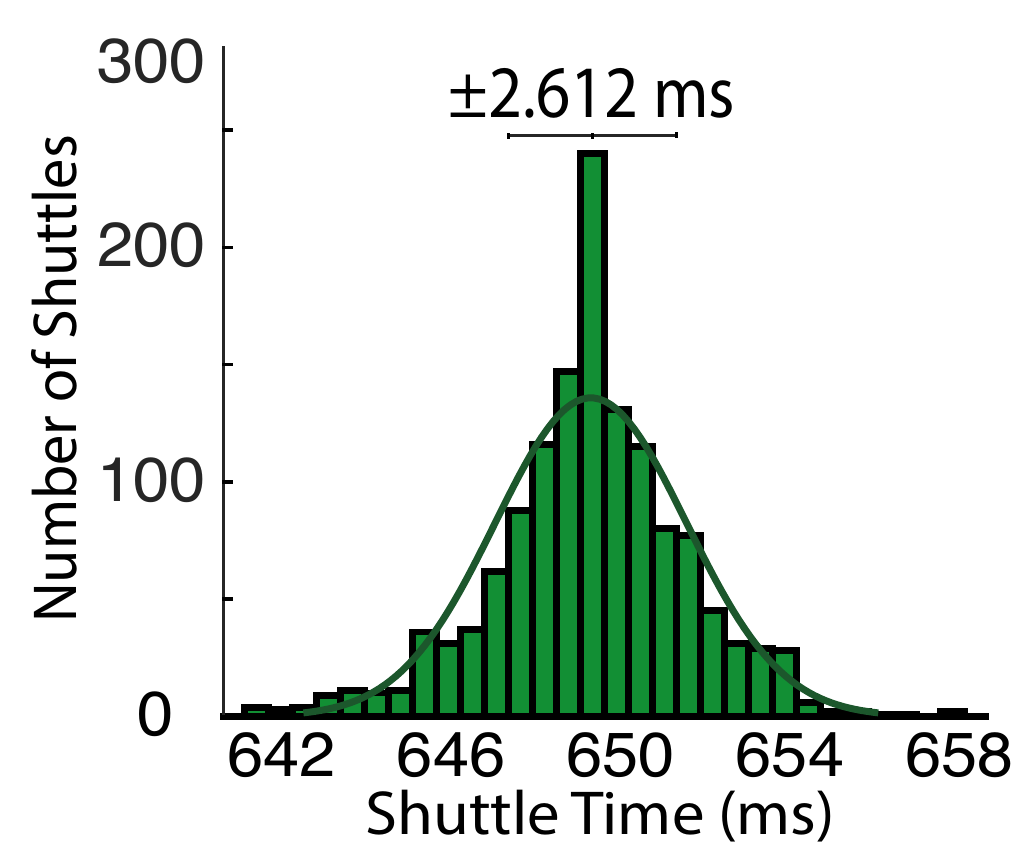}
  \caption{\textbf{Sample shuttling repeatability.} Shuttler operation (1400 runs) between polarization ($\sim$30mT) and detection (7T) locations, distance of approximately 928mm depending on sample holder inserted. Samples are pressure held onto hollow carbon fiber rod along the center of the magnet bore and shuttled using a mechanical actuator activated by synchronized pulse trigger. This demonstrates high stability for repeated experiments, with average travel times of 648$\pm$0.6ms.}
  \zfl{histogram}
\end{figure}

\begin{figure}[t]
  \centering
  \includegraphics[width=0.4\textwidth]{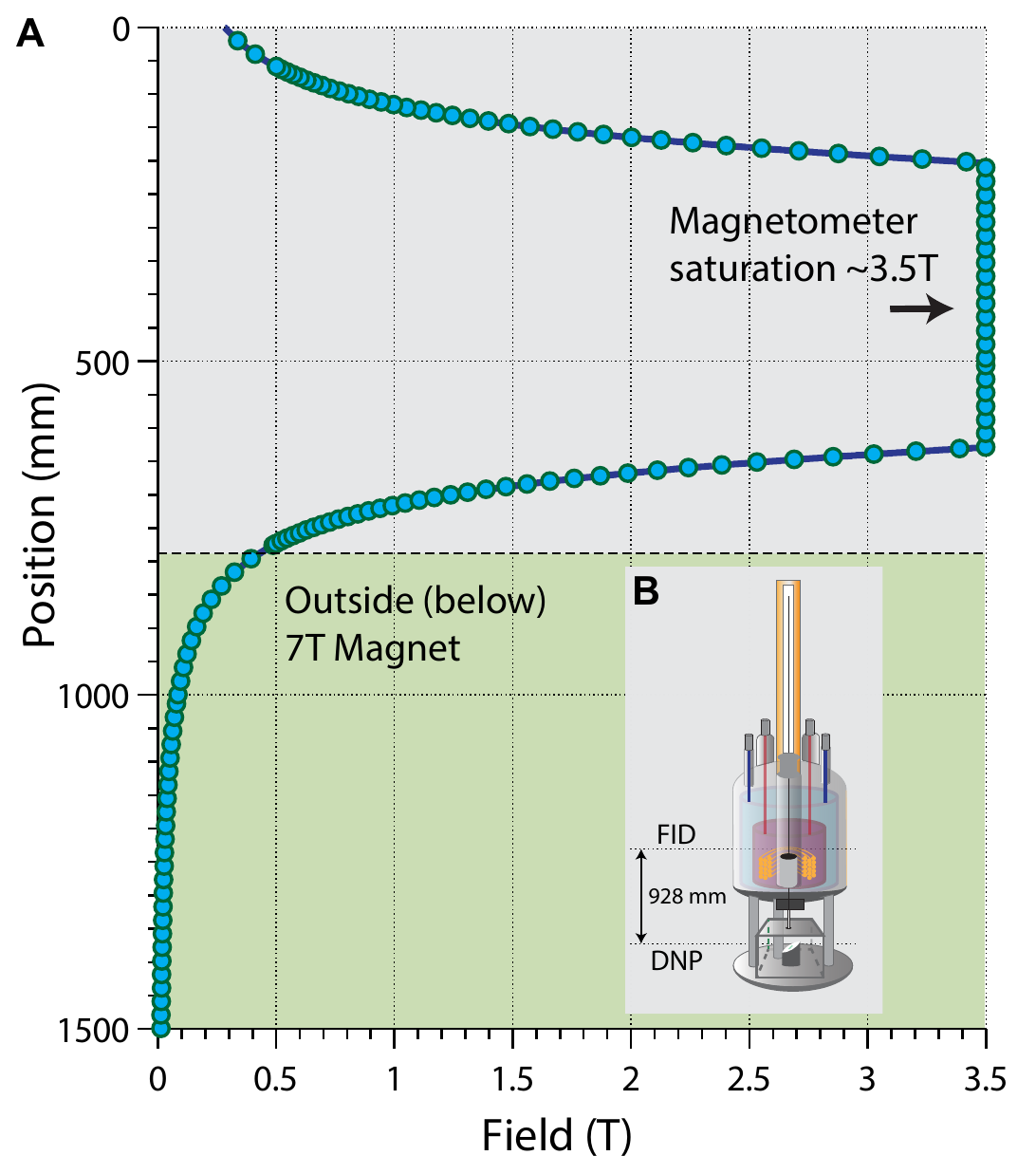}
  \caption{\textbf{Field map}  (A) Measurement of longitudinal (z axis) magnetic field over full field cycler range using a sensitive magnetometer. Data points were attained by shuttling magnetic field probe through center of the magnetic bore while held within the hollow carbon fiber shuttling rod, limiting accuracy to the 50$\mu$m precision of the actuator. Position of magnet entrance is shown to demonstrate fringe field profile. Due to magnetometer constraints, high field measurements saturate at 3.5T. (B) Polarization is generated $\sim$928mm from the NMR coil, depending on the sample holder. This range can be traveled in sub-second speeds (see \zfr{field_cycler}), allowing fast transport of hyperpolarized diamonds from low fields below to center of magnet with minimal relaxation loss.}
  \zfl{field_map}
\end{figure}

\section{Field Cycling}
$T_1$ noise spectroscopy relies on our ability to rapidly vary the magnetic field experienced by a test sample using a homemade shuttling system built over a 7T superconducting magnet~\cite{Ajoyinstrument18}. Samples are held in an NMR tube (Wilman 8mm OD, 1mm thickness) (see\zfr{field_cycler}D) and pressure-fastened from below the magnet onto a lightweight, carbon fiber shuttling rod (Rock West composites). Using a high precision (50$\mu$m) conveyor actuator stage (Parker HMRB08) (see \zfr{field_cycler}B), we are able to repeatedly and consistently shuttle from low fields ($\sim$30 mT) below the magnet for polarization to high fields (7T) within the magnet for NMR detection at sub-second speeds ($<$700ms). The instrument is interfaced with a low-cost hyperpolarizer (See ~\cite{Ajoy18pol} for details) %cite hyperpolarizer paper
 , allowing generation and detection of bulk nuclear polarization. Because the average shuttling time is small compared to the nuclear $T_1$ lifetimes (see \zfr{histogram}) -- particularly at fields above 100mT -- our resulting NMR signals are recorded with minimal loss in enhancement. High precision shuttling allowed for the measurement of a full z-direction field map (see \zfr{field_map}) %Instrument paper
 ), where the field was measured as a function of position using an axial Hall probe for fields less than 3.5T. To accommodate the fast shuttling technique, the conventional NMR probe was modified to be hollow, allowing for shuttling through the probe to low magnetic fields below the magnet. Custom made ``printed'' coils (see~\cite{coilvideo2}) are employed for direct inductive detection of the NMR signals ~\cite{Ajoyinstrument18}.

\begin{figure}[t]
  \centering
  \includegraphics[width=0.5\textwidth]{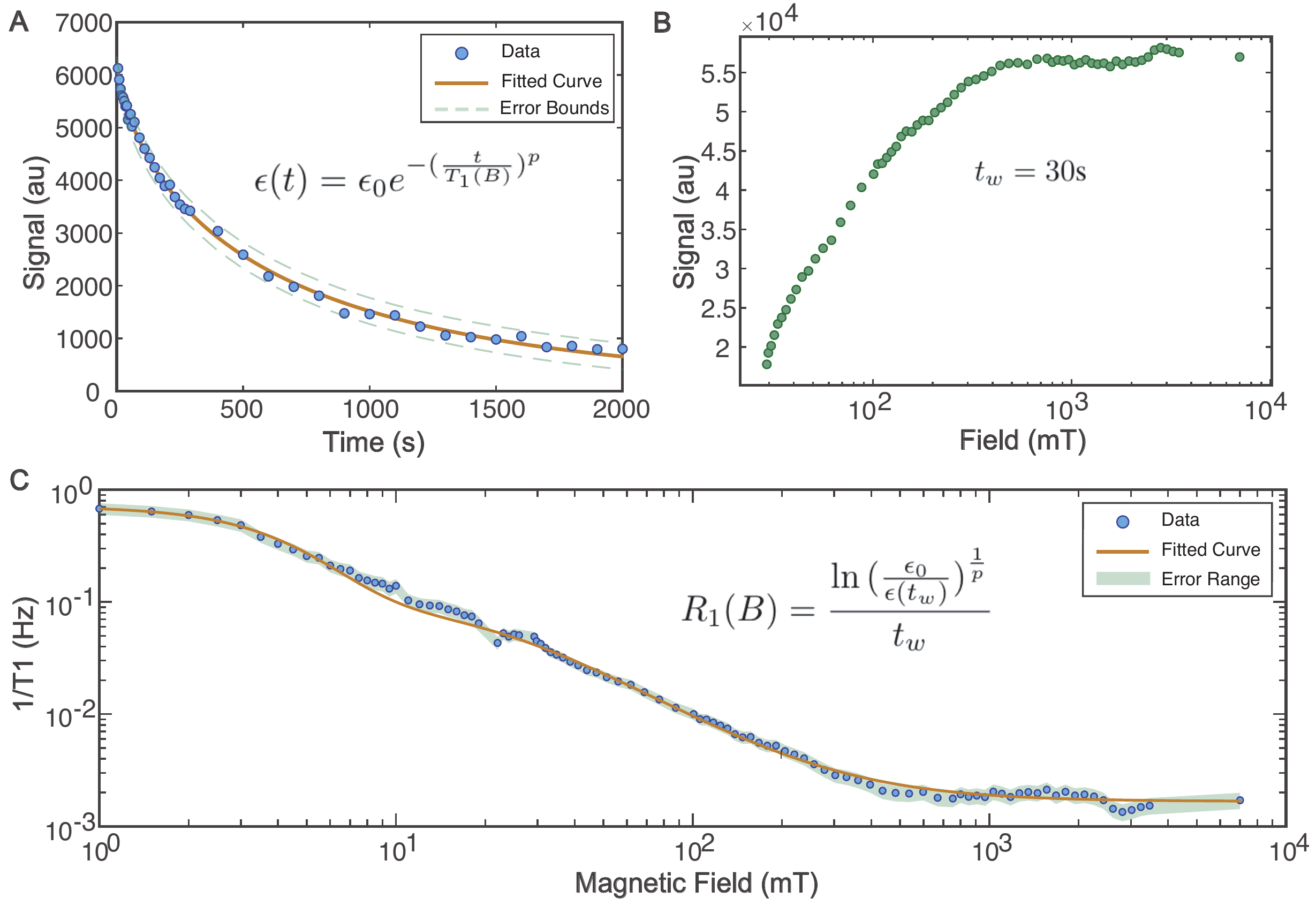}
  \caption{\T{Data processing.} (A) Spin polarization decay curves are acquired by repeated hyperpolarization of the diamond sample followed by time-dependent relaxation at a given field. By varying wait time and measuring the resulting NMR signal, relaxation parameters at this field can be estimated by fitting the data to a stretched exponential function. Because the relaxation rate equation incorporates a phenomenological stretch factor to account for T1 heterogeneity at different fields, decay experiments are done at varied fields and the fitted parameters are used for different field regimes. (B) enhancement data is also taken at varying fields with wait time kept constant, providing a 1D slice of the relaxation dynamics at wide field ranges. To maximize signal contrast the wait times are dynamically adjusted to account for different T1 regimes. (C) Using the two previous experiments, a relaxation field map is constructed using the estimated rate equation parameters and the 1D enhancement data. Errors result from the quality of the decay curve fits and inaccuracies in the measured magnetic field.}
  \zfl{data_processing}
\end{figure}

\section{Data Processing}

%%% DECAY CURVES, STRETCHED EXPONENTIAL %%%

\subsection{Fit models}

Nuclear $T_1$ at a given magnetic field is determined by measuring the decay of NMR signal $\vxe(t)$ with respect to time $t$ spent decaying at that field. By measuring the change in signal over various times, relaxation decay curves are determined, and $T_1(B)$ estimated. We find that all the data can be fit to a stretched exponential of the form (see \zfr{data_processing}A),
\beq
\vxe(t) =\vxe_0e^{-(\frac{t}{T_1(B)})^p}\:,
\eeq
where $p \in (0, 1]$ is a stretch factor \cite{Jarmola12}, and $\epsilon_0$ represents the bare signal enhancement obtained from DNP and assuming no loss during shuttling. For certain samples, such as the 10\% $\Cs$ sample in \zfr{ten}C, we observe that $p\app 1$, while for most samples with low $\Cs$ enrichment (including at natural abundance), $p\in (0.5,1)$. We ascribe this stretch factor to be arising from spin diffusion of the inhomogeneous polarization in the lattice that is driven by the DNP process.

%%% RELAXATION PROFILES, TSALLIS DISTRIBUTION %%%

%%Figure S3. Data processing. (A) Spin polarization decay curves are acquired by repeated hyperpolarization of the diamond sample followed by time-dependent relaxation at a given field. By varying wait time and measuring the resulting NMR signal, relaxation parameters at this field can be estimated by fitting the data to a stretched exponential function. Because the relaxation rate equation incorporates a phenomenological stretch factor to account for T1 heterogeneity at different fields, decay experiments are done at varied fields and the fitted parameters are used for different field regimes. (B) enhancement data is also taken at varying fields with wait time kept constant, providing a 1D slice of the relaxation dynamics at wide field ranges. To maximize signal contrast the wait times are dynamically adjusted to account for different T1 regimes. (C) Using the two previous experiments, a relaxation field map is constructed using the estimated rate equation parameters and the 1D enhancement data. Errors result from the quality of the decay curve fits and inaccuracies in the measured magnetic field.

By measuring the relaxation rate $R_1(B) = 1/T_1(B)$ over a range of magnetic fields allowed by the field cycler, a relaxation field map $R_1(B)$ can be obtained, as shown in \zfr{ten}B. These relaxation profiles are then fit to a sum of two Tsallis distributions [36], a generalization of Gaussian and Lorentzian functions that allows greater flexibility in representing the relaxation rate as a function of field. Additionally our model assumes a constant offset to account for the saturation of the relaxation rate at high field, with functional form of a single Tsallian with respect to field $B$,
\beq
R_1(B)  =  C_1\left[1 + (2^{q-1} - 1)\left(\frac{B}{C_2}\right)^2\right]^{-\frac{1}{q-1}} + C_3
\eeq
where fitting parameters $C_1, C_2, C_3$ describe the amplitude, width and vertical offset of the function respectively, and  $q$ regulates the effective contribution of the function's tail to the overall area under the function, with pointwise limits $q=1$ and $q=2$ denoting Gaussian and Lorentzian functions respectively. Originally the fitting models were limited to either Lorentzian/Gaussian lineshapes, and the model was susceptible to deviate from the experimental relaxation estimates at high field. By allowing variation of the parameter $q$, qualitatively better fits to the relaxation profiles can be found and analyzed in relation to one another. 

\subsection{Accelerated data collection strategy}
Due to long relaxation times at high field, occasionally approaching $\sim$20 minutes, production of enhancement decay data at an array of magnetic fields is time-intensive. In order to hasten measurement times, and to obtain a denser map of nuclear $T_1$ estimates at a large number ($\sim$100) of field points (for example in \zfr{one_percent}), we created an accelerated (yet approximate) measurement strategy that we now detail. After hyperpolarization and subsequent transfer to the field of interest, the signal $\vxe(t_w)$ after some fixed wait time $t_w$ (typically 30s) at a certain field is measured (see \zfr{data_processing}B). Because the sample decays for the same time at each field, this set of enhancement values provides a hint as to the relaxation mechanisms throughout the full field range. To estimate $T_1$ from this data, however, requires knowledge of the enhancement generated before relaxation begins. To estimate this quantity, hereafter referred to as $\vxe_0$, decay curves are experimentally acquired at certain fields using several averages per experiment, ensuring low error when fitting this curve to a stretched exponential model. Using the fit parameters $T_1$ and $p$, $\vxe_0$ can then be estimated as
\beq
\vxe_0 = \vxe(t_w)e^{\lb\frac{t_w}{T_1}\rb^p}
\eeq
This estimate allows us to reconstruct the relaxation rate at each field for which enhancement measurements were acquired. By reordering the relaxation equation, the estimate of $R_1$ at field $B$ becomes
\beq
R_1(B) = \frac{\ln{\lsb\frac{\vxe_0 }{ \vxe(t_w)}\rsb}^{\frac{1}{p}}}{t_w}
\eeq

The quality of this reconstruction is improved by doing multiple decay curve experiments at varying fields so that the appropriate stretch factor $p$ can be determined for different field regimes. For the two natural abundance $\Cs$ samples in \zfr{one_percent} we used decay curve data at fields of 20mT, 35mT, 150mT, and 7T for the relaxation field reconstructions, with stretch factors $p\app$0.75 at lower fields and $p\app$1 at high fields. For the enriched samples in \zfr{enriched}, the approximation method for relaxation data was used for the 3\% sample whereas the other sample data was acquired using the 2D decay curve procedure. 

In certain cases, especially for the ultralow field data in \zfr{one_percent}, rather than using a constant decay time $t_w$ for all points, the sensitivity of the decayed enhancement readings is maximized by using \I{dynamically} varied wait times $t_w$ at different fields; the loss in enhancement then becomes approximately 50\% of the initial polarization value. This process mitigates errors in the measured enhancement values by creating sufficient contrast between the initial and decayed enhancement values, without excessively diminishing the signal relative to the noise.

Let us now quantify the time savings resulting from this data collection strategy. By removing the need to explicitly plot the signal decay over time at every magnetic field point, the effective dimensionality of our $T_1(B)$ measurement process is reduced, which allows determination of $T_1$ at a large number of field points rapidly. To develop an intuition for the accelerated in the averaging time gained as a result, we assume an even sampling of the signal decay, in time increments $\Delta t$ across $n$ steps. To obtain estimates of $T_1$ at $N$ field values, this would require at the very least a total time $t_{2D} = N \Delta t \sum_i^n i =  N \Delta t \frac{n(n+1)}{2}$. While employing the accelerated 1D measurement strategy in contrast, signal enhancement is measured after a fixed wait time $t_w$ at each field. These measurements are obtained at all $N$ field points, after sampling with high accuracy the signal decay curves at $N_{d}$ overlapping fields to construct estimates of the initial enhancement and stretch factor at varied fields. The experiment would therefore expend a minimum time of $t_{\text{1D}}  = N t_\text{w} + N_{d} \Delta t n (n+1)/2 $. This measurement strategy incurs a theoretical time gain of $\frac{t_\text{2D}}{t_\text{1D}}  = \frac{N\Delta t n(n+1)}{2Nt_w + N_{d}\Delta t n(n+1)}$, with the simplifying assumption that zero time is spent moving between fields as well as during signal detection.  To demonstrate the possible time gains of this method, assume signal decay measurements at $\Delta t=10$s increments for a total of $n=40$ points in time, across $N=100$ field points. This may then be compared to the accelerated 1D measurement strategy, with signal enhancement measurements after a fixed hyperpolarization time of $t_w$ = 30s at each field. If $N_d = 4$ decay curves are used to estimate the relevant relaxation properties at four separate fields, the time gain of the 1D strategy is $\frac{t_\text{2D}}{t_\text{1D}} \approx 23$.

\subsection{Error estimates}
%%% Errors %%%
Let us now outline the error estimation in the $T_1(B)$ data. The primary sources of error come from the tightness of the decay curve fits to estimate $\vxe_0$ and $p$ at different fields, the shot-to-shot error in the measured enhancement, and the error in the wait time spent relaxing at a given field. Because of the high averaging done to generate relaxation decay curves, the error in $\vxe_0$ and $p$, taken from the fitting function confidence intervals, is very small $\app$1\%. To account for variation in the relaxation wait time, the two methods used for placing the sample at a given field are considered. To access high field points the sample is shuttled into the magnet and allowed to wait a set time, and the error in this process arises from the shuttling time. Because the field cycler can shuttle the sample over the maximum field range in less than 1 second, the shuttling error is approximated as 2s.  To access the low field regime, a bidirectional Helmholtz coil was assembled within the hyperpolarizer which is aligned with the field produced by the superconducting magnet in the $+z$ direction. This allows us to probe fields lower than what is covered by the field cycler. At the polarization location and with no current driven through the coil, the 7T magnet produces a field of 20.8mT, but fields as low as 1mT and even further can be attained with use of the coil. To account for the build-up of magnetic field due to the coil, we attribute an error of 2s to all points found by this process. In combining both shuttled and coil-generated field points there was a constant offset of 15mT added to all shuttled field points to make the curves consistent with the low field relaxation rate points. 

\begin{figure*}[t]
  \centering
  \includegraphics[width=0.9\textwidth]{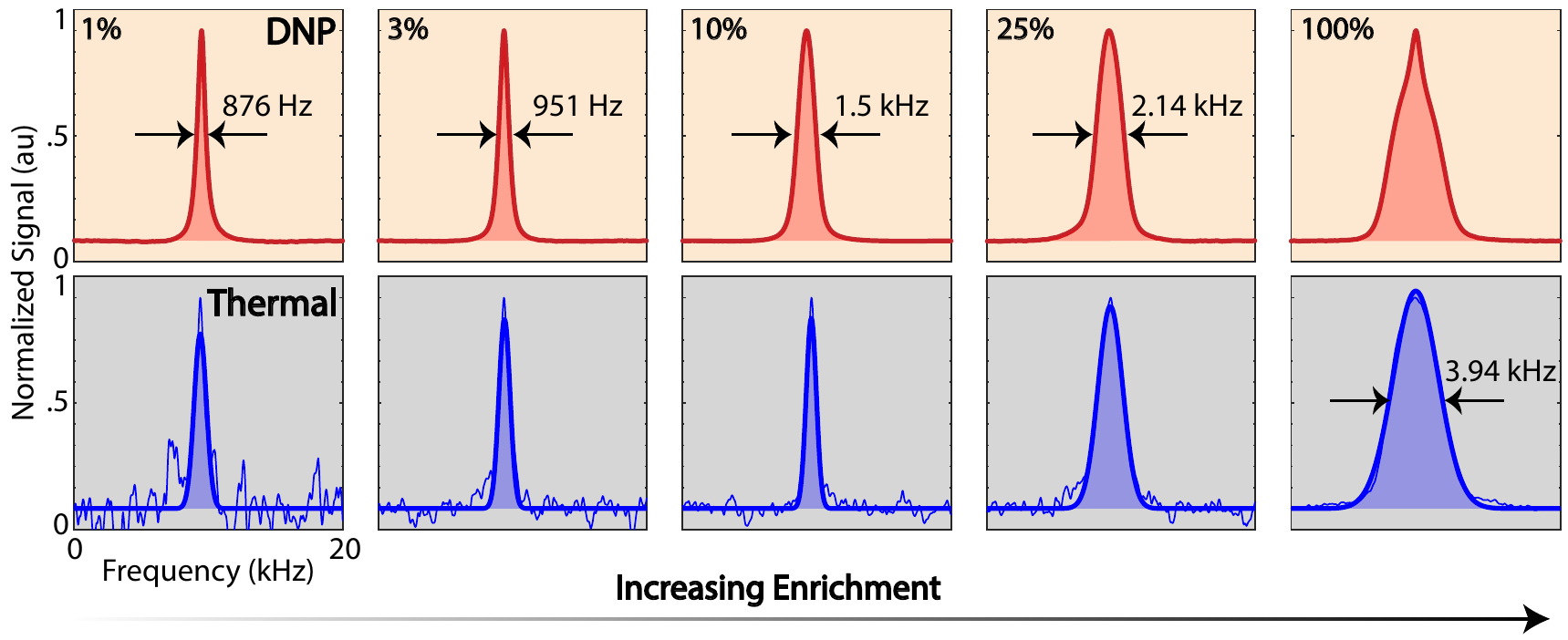}
 \caption{\textbf{Comparison of DNP and thermal $\Cs$ lineshapes.} Panels indicate lineshapes under (A) hyperpolarization carried out at low field (1-30mT) and (B) 7T thermal polarization. DNP is excited from the optically polarized NV centers which are $\app$1ppm in all samples.  For the 100\% sample, we ascribe the broad and narrow components of the lineshapes (dashed lines) as being spins close and further away from the NV centers respectively. The scaling of the experimental linewidths matches our predictions from theory (see \zfr{carbon}C).}
\zfl{lineshapes}
\end{figure*}

\section{Model For Hyperpolarized Relaxometry}
We now provide more details of the model employed to capture the relaxation mechanisms probed by our experiments.   We had identified from the experiments three relaxation channels that are operational at different field regimes, driven respectively by \I{(i)} couplings of the $\Cs$ nuclei to pairs (or generally the reservoir) of P1 center, \I{(ii)} individual P1 or NV centers, and \I{(iii)} due to spin-diffusion effects within the $\Cs$ reservoir. In this section, we detail lattice calculations that allow the estimation of the spectral densities in each of these cases.

Consider again the three disjoint spin reservoirs in the diamond lattice, the electron spin reservoir of NV centers, electron reservoir of  substitutional-nitrogen (P1 centers), and the $\Cs$ nuclear spin reservoir. They are centered respectively at frequencies $\xo_{\R{NV}}\app[(\xD\pm \xg_e\Br\cos\xt_{\R{NV}})^2 + (\xg_e\Br\sin\xt_{\R{NV}})^2]^{1/2}$, $\xo_{e}\app[(\xg_e\Br + m_IA_{\pll}^{\R{P1}}\cos\xt_{\R{P1}})^2 + (m_IA_{\pp}^{\R{P1}}\sin\xt_{\R{P1}})^2]^{1/2}$ and the nuclear Larmor frequency $\xo_L=\xg_n\Br$; where $\xt_{\R{NV}},\xt_{\R{P1}}$ are angles of the NV(P1) axes to the field, $A_{\pll}^{\R{P1}}\app$114MHz, $A_{\pp}^{\R{P1}}\app$86MHz are the hyperfine field of the P1 center to its host $\Ns$ nuclear spin, $m_I=\{-1,0,1\}$ is the $\Ns$ manifold, $\xD$=2.87GHz is the NV center zero field splitting, and $\xg_e=28$MHz/G and $\xg_n=1.07$kHz/G are the electronic and nuclear gyromagnetic ratios.

\subsection{Lattice estimates for electron reservoir}
\zsl{P1}
In order to determine the relaxation in behavior \zr{electron_relax} quantitatively, let us determine typical inter-spin couplings and distances for the electron reservoir from lattice concentrations. First, for the electronic spins, given the relatively low concentrations, and the fact that the lattice is populated independently and randomly, we make a Poisson approximation following Ref.~\cite{Reynhardt03a}. An estimate for the \I{typical} inter-spin distance $\expec{r_e}$ is obtained by determining the distance at which the probability of finding zero particles is $\frac{1}{2}$. Given the lattice spacing in diamond $a$=0.35nm, and the fact that there are four atoms per unit cell, we can estimate the electronic concentration in inverse volume units as, $N_e=(4\zt 10^{-6}P_e)/a^3$[m$^{-3}$]. Then from the Poisson approximation 
$
\expec{r_e} =\lb 3/4\pi \ln 2\rb^{1/3}N_e^{-1/3}
$we obtain, for instance, $\expec{r_{\R{NV}}}=$12.12nm and $\expec{r_{\R{P1}}}=$2.61nm, where we have assumed concentrations of 1ppm and 100ppm respectively.

The inter-spin distances now allow us to calculate the second moment of the electronic spectra, which are reflective of the mean inter-spin couplings. Following Abragam~\cite{Abragam61}, we have 
\beq
M_{2e} = \fr{9}{20} (g\mu_B)^2\fr{1}{\expec{r_e}^6},
\eeq
where $g\app 2$ is the electron g-factor, and $\mu_B=9.27\zt10^{-21}$erg/G the Bohr magneton in cgs units. Substituting this leads to, $M_{2e} =43.65P_e^2$ [mG$^2$], and allows us to estimate the electronic line width, 
$\xD f_e=\expec{d_{ee}}\app \xg_e\sq{\fr{8}{\pi}}\sq{M_{2e}}\:$[Hz]$\app$10.5$P_e$[mG], that scales approximately linearly with electron concentration $P_e$. Here we have assumed a Lorentzian lineshape and quantified the linewidth from the first derivative~\cite{Reynhardt03a}. Typical values are $\xD f_{\R{NV}}$=29.52kHz and $\xD f_{\R{P1}}$=2.95MHz at 1ppm and 100ppm concentrations respectively.

Let us now estimate the effective hyperfine interaction from the P1 centers to the $\Cs$ reservoir. Our estimate can be accomplished by sitting on a P1 spin, and evaluating the mean perpendicular hyperfine coupling that contributes to the spin flipping noise, $\expec{A_{zx}} = \lsb\expec{A_{zx}^2}\rsb^{1/2}$, where we setup the second moment sum,
\beq
\expec{A_{zx}^2} = \fr{1}{N}\lsb \fr{\mu_0}{4\pi}\xg_e\xg_n\hbar\rsb^2\sum_{j}\fr{(3\sin\xt_j\cos\xt_j)^2}{r_j^6}
\zl{Azx}
\eeq
where $N$ is the total number of $\Cs$ spins for every P1 center and $\xt_j$ is the angle between the P1-$\Cs$ axis and the magnetic field.  Numerically the factor $\fr{\mu_0}{4\pi}\xg_e\xg_n\hbar=$19.79[kHz (nm)$^3$]. For simplicity, we can approximate the sum by an integral, and including the density of $\Cs$ spins $N_C=0.92\eta$ spins/nm$^3$ (see \zfr{carbon}B), where $\eta$ is the $\Cs$ enrichment level,
\beq
\expec{A_{zx}^2} = \lb \fr{\mu_0}{4\pi}\xg_e\xg_n\hbar\rb^2\fr{N_C(2\pi)}{N_C\mV}\int_{r_0}^{\expec{r_e}}\int_0^{\pi/2}\fr{(9\sin^3\xt\cos^2\xt)}{r^6} r^2drd\xt\non
\eeq
where $\mV=\fr{4\pi}{3}\expec{r_e}^3$ corresponds to the volume of spins considered. We have assumed that the ``sphere of influence'' of a particular P1 spin notionally extends to the mean distance between neighboring P1 centers, for instance $\expec{r_e}=$5.62nm for $P_e$=10ppm. The integral lower limit is set by the requirement that the hyperfine shift of the $\Cs$ nuclei is within the detected NMR linewidth  $\xD f_{\R{det}}\app$2kHz. Then, $r_0=[19.79/(\xD f_{\R{det}})]^{1/3}\app$2.15nm. In principle, $r_0$ goes to quantify a ``barrier'' around around each P1 center, wherein the hyperfine interactions prevent the $\Cs$ nuclei from being directly observable in our relaxometry experiments. The angle part of the integral evaluates to $6/5$, and effectively therefore,
\beq
\expec{A_{zx}^2} = \lb \fr{\mu_0}{4\pi}\xg_e\xg_n\hbar\rb^2 \fr{6}{5} \fr{1}{\expec{r_e}^3}\lb \fr{1}{r_0^3} - \fr{1}{\expec{r_e}^3}\rb
\zl{hyp_eqn}
\eeq
For instance, for the two natural abundance single crystal samples that we considered in the \zfr{one_percent} of the main paper with P1 concentration 17ppm and 48ppm, we have $\expec{r_e}=$ 4.8nm and 3.39nm respectively, giving rise to the effective P1-$\Cs$ hyperfine interaction $\expec{A_{zx}^2}\app$ 0.39[(kHz)$^2$] and $\expec{A_{zx}^2}\app$0.45[(kHz)$^2$] respectively. The simple model predicts that the effective hyperfine coupling increases slowly with the electron concentration $P_e$,  that the electron spectral density width $\expec{d_{ee}}\propto P_e$. It also shows that the electron spectral density is independent of $\Cs$ enrichment $\eta$ to first order. The zero-field relaxation rates stemming from this coupled-electron mechanism can now be calculated as $R_1(0) = \expec{A_{zx}^2}/\expec{d_{ee}}\app$ 777[s$^{-1}$] and 317.5[s$^{-1}$]. This matches our expectation for the order of magnitude of the zero field rate since we expect that the $\Cs$ relaxation time $T_{1n}$ matches that of the electron $T_{1e}\app$1ms.

In order to validate the conclusions from this simple model, we perform an alternative numerical estimation of $\expec{A_{zx}^2}=\lsb \fr{1}{N}\sum_{j\in \xD f_{\R{det}}}\expec{A^2_{zx,j}}\rsb$ within the detection barrier directly from the diamond lattice (see \zfr{carbon}F and \zsr{NV}). We obtain $\expec{A_{zx}^2}=$2[(kHz)$^2$] and 2.26[(kHz)$^2$] for the $P_e$=17ppm and 48ppm samples respectively, in close and quantitative agreement with the values predicted from \zr{hyp_eqn} (considering the approximations made in the analysis). Numerics also confirm that the hyperfine values $\expec{A_{zx}^2}$ are independent of enrichment $\eta$ (see \zfr{carbon}F) in agreement with the experimental data.

\begin{figure*}[t]
  \centering
  \includegraphics[width=\textwidth]{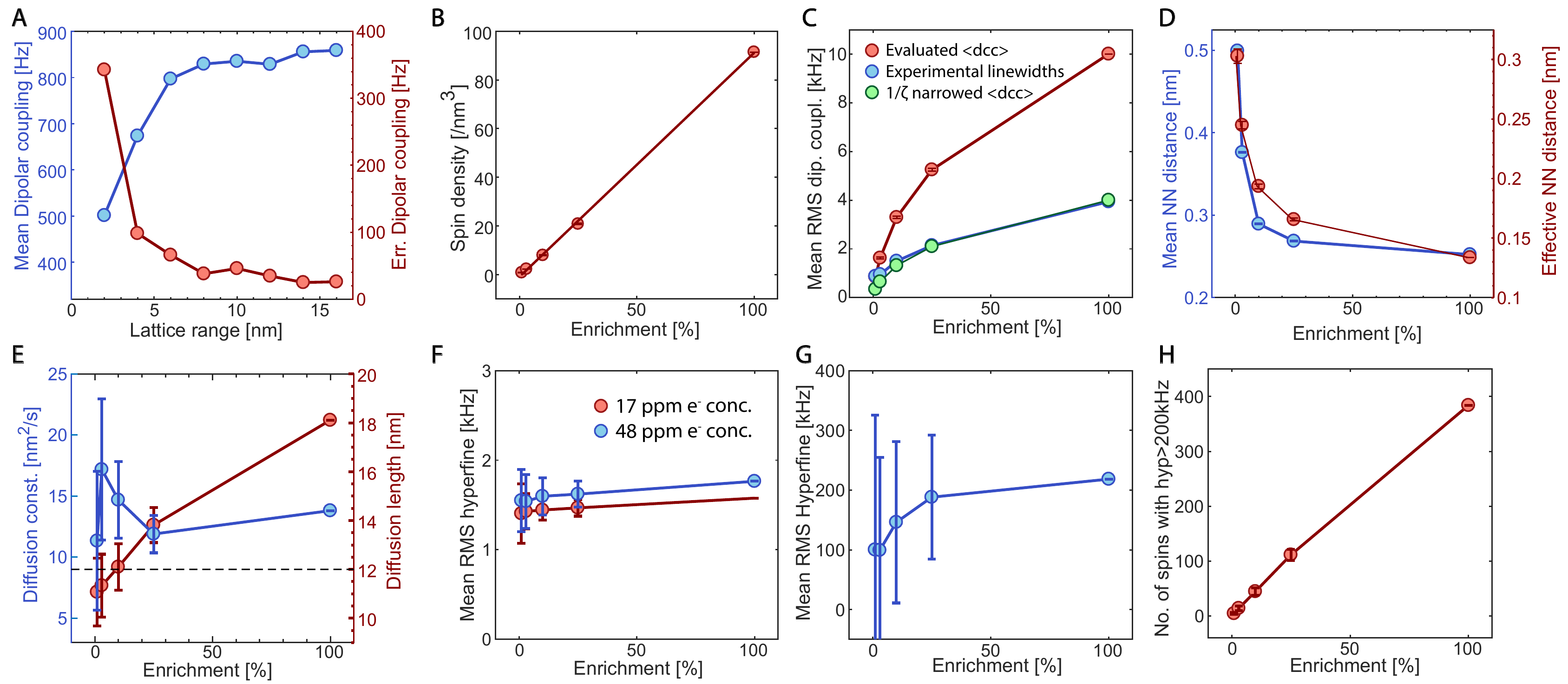}
  \caption{\T{Calculated interspin parameters pertaining to $\Cs$ and NV reservoirs as a function of lattice enrichment $\eta$.} (A) \I{Convergence of numerical estimates} is representatively illustrated by plotting the mean $\Cs$ dipolar coupling $\expec{d_{\CC}}^{\ell}$ and the residual $\xe(\ell)$ as a function of considered lattice size $\ell$. We evaluated here the case of a 1\% enriched diamond single crystal. We observe good convergence beyond a lattice size of about 10 nm. (B) \I{Spin density} of $\Cs$  nuclei shows, as expected, very close to linear dependence with $\eta$. Solid line is a linear fit, whose slope returns the lattice spin density $\app0.92\eta$spins/nm$^3$. (C) \I{Effective inter-nuclear dipolar coupling} $\expec{d_{\CC}}$ evaluated from second moment (red line). Blue points show the experimentally obtained linewidths. Green line indicates $\fr{1}{\zeta}\expec{d_{\CC}}$ with broadening factor $\zeta=2.5$, and shows a good numerical agreement with experimental data. (D) \I{Mean inter-spin distance} $\expec{r_n}$ between lattice $\Cs$ nuclei in evaluated from the RMS dipolar coupling (red points) and from effective nearest-neighbor lattice distances (blue points). The two estimates show a good match, with the inter-spin distance falling approximately as $\eta^{1/3}$. (E) \I{Diffusion constant and diffusion length} numerically estimated with lattice enrichment. Here we employed experimentally obtained values of $\Cs$ T1. Dashed line indicates the mean inter-electron distance between NV centers at 1ppm concentration, indicating that spin diffusion can homogeneously spread polarization in the lattice almost independent of $\Cs$ enrichment. (F) \I{Effective hyperfine coupling} $\expec{A_{zx}^{\obs}}$ to P1 centers in case of single crystal samples with 17ppm (red points) and 48ppm (blue points) electron concentration. Results indicate that $\expec{A_{zx}^{\obs}}$ is independent of $\Cs$ enrichment $\eta$. 
(G) \I{Estimates of mean RMS NV-$\Cs$ hyperfine interaction} $\expec{A_{\NV}}$ with lattice enrichment. 
(H) \I{Estimation of directly participating $\Cs$ nuclei} in the DNP process, defined as those nuclei for which the hyperfine coupling to the closest NV center is greater than 200 kHz. We obtain an approximately linear increase with enrichment. Error bars in all panels are numerically estimated from standard deviation of lattice parameter distributions over several realizations of the lattice configuration.
}
  \zfl{carbon}
\end{figure*}

\subsection{Lattice estimates for $\Cs$ reservoir}
\zl{C13}
In contrast, since the $\Cs$ reservoir has a much larger spin density, especially at high enrichment levels, we will estimate the interspin distances $\expec{r_n}$ and couplings $\xD f_n$ numerically. The experimentally obtained $\Cs$ lineshapes and resulting linewidths for all the samples considered are shown in \zfr{lineshapes}. We begin by first setting up a diamond lattice numerically and populating the $\Cs$ spins with enrichment level set by $\eta$. The numerical calculation is tractable since only small lattice sizes typically under $\ell$=10nm are sufficient to ensure convergence of the various dipolar parameters (see \zfr{carbon}A). To a good approximation, we determine the spin density of the $\Cs$ nuclei to be $N_C=0.92\eta$ spins/nm$^3$ (see \zfr{carbon}B). Next, in order to determine the nuclear dipolar linewidths, we consider the secular dipolar interaction between two nuclear spins $j$ and $k$ in lattice, 
\beq
d_{jk}^{\CC} = \fr{\mu_0}{4\pi}\hbar\xg_n^2(3\cos^2\xt_{jk}-1)\fr{1}{r_{jk}^3}
\zl{djk}
\eeq
where $\xt_{jk}=\cos^{-1}\lb \fr{\T{r}_{jk}\cdot\Bpol}{r_{jk}\Bp}\rb$ is the angle between the inter-nuclear vector and the direction of the magnetic field. In the numerical simulations we will consider, we evaluate the case of single crystal samples placed flat, i.e. with $\Bpol\pll$ [001] crystal axis. As a result, for $\Cs$ spins on adjacent lattice sites, $\xt_{jk}=$54.7$^{\circ}$ is the magic angle and $d_{jk}^{\CC}=0$. We note that \zr{djk} is a good approximation even \I{during} the hyperpolarization process. Indeed, although hyperpolarization is performed in the regime where the nuclear Larmor frequency $\xo_L$ is smaller than the hyperfine interaction $A$ to the NV center, the hyperfine field is only transiently �on� during the microwave sweep. Given the fact that the NV center is a spin-1 electron, there is no hyperfine field applied to the nuclei when the NV is optically pumped to the $m_s=0$ spin state. Indeed this constitutes the majority of time period of the DNP process.

We now evaluate the \I{effective} mean dipolar coupling $\expec{d_{\CC}}$ between the nuclei from the second moment, 
\beq
\expec{d_{\CC}} = \fr{1}{N}\sum_j\lsb\sum_k \lb \fr{\mu_0}{4\pi}\hbar\xg_n^2(3\cos^2\xt_{jk}-1)\rb^2\fr{1}{r_{jk}^6}\rsb^{1/2}\:,
\zl{dipC}
\eeq
where $N=N_C\ell^3$ refers to the number of $\Cs$ spins in the lattice, and for the convergence, we assign for simplicity, $1/r_{jj}$=0. This simply allows us to sum over all the spins $j$ in the lattice. In practice, we evaluate the parameter $\expec{d_{\CC}}$ in \zr{dipC} over several ($\app$ 20) realizations of the lattice and take an ensemble average (see \zfr{carbon}C). We report an effective error bar from the standard deviation of this distribution. The fidelity of the obtained results is evaluated by testing the convergence $\xe(\ell)=\|\expec{d_{\CC}}^{\ell+1} -\expec{d_{\CC}}^{\ell}\|$, where the $(\ell+1)$ superscript indicates a lattice expanded by 1nm. As is evident in the representative example for $\eta=$1.1\% displayed in \zfr{carbon}A, we find good convergence ($\xe\rt 0$) for $\ell\app$14nm, corresponding to about 2500 lattice $\Cs$ nuclei.

It is instructive to now compare the estimated values with the experimentally determined nuclear linewidths $\xD f_n(\eta)$ measured at 7T (see \zfr{lineshapes} and blue points in \zfr{carbon}C). The scaling (solid line in  \zfr{carbon}C) of the experimental data $\sim\eta^{1/2}$ matches closely with the estimated result through \zr{dipC} (see red line in \zfr{carbon}C). However we find that the numerical value overestimates the linewidth by an additional broadening factor $\xz\app 2.5$. The green points show a close match between experimental values and numerically evaluated $\fr{1}{\xz}\expec{d_{\CC}}$.

% We note that the theoretical values in \zr{dipC} provide an underestimate since they assume that the $\Cs$ linewidth arises solely from dipolar inter-spin interactions, and ignore hyperfine couplings to the NV center and P1 center reservoirs. Since the electron spins are thermally polarized to $\app$3\% at 7T, this effect is non-negligible. Assigning this extra broadening to be an additional factor $\xz$, we find very close match (see \zfr{carbon}C) between the experimental values and the numerical evaluated ones $\xz\expec{d_{\CC}}$, with $\xz\app$2.5.

This effective coupling now allows us to estimate the mean inter-spin distance $\expec{r_n}$ as a function of $\Cs$ enrichment (see \zfr{carbon}D), 
\beq
\expec{r_n} = \lsb \fr{2\expec{d_{\CC}}}{\fr{\mu_0}{4\pi}\xg_n^2\hbar}\rsb^{-1/3}
\eeq
We find a scaling $\sim\eta^{-1/6}$ (red line in \zfr{carbon}D). It is also interesting to compare these values to those alternatively evaluated directly from the lattice (blue points in \zfr{carbon}C). For this, we rely on the fact that the $\expec{r_n}$ distances largely reflect the nearest-neighbor (NN) spin distances. We define the NN spin (say $k$) to the spin $j$ as the one which has the dipolar coupling $d_{jk}$ is maximal. Now for every spin $j$ in the lattice, we determine the nearest neighbor inter-spin distance $R_j=\labs r_{jk}^{\R{NN}}\rabs$, and construct a row matrix, $\T{R}=\lcb R_j\rcb$, with $j^{\R{th}}$ element $R_j$. Finally,  repeating and contacentating this row matrix for several realizations of the lattice, we finally estimate $\expec{r_n} = \expec{ \T{R}}$ for the i$^{\R{th}}$ realization of the lattice. The comparison between these two metrics is demonstrated in \zfr{carbon}D), and show reasonably good agreement.

These inter-spin distances and the coupling values allow us to estimate the spin diffusion coefficient $D(\eta)$ as a function of lattice enrichment (see \zfr{carbon}E). This quantifies the spread of polarization away from directly polarized $\Cs$ nuclei, and also serves as a means to quantify the �homogenization� of polarization in the lattice. Following Ref.~\cite{Hayashi08}, we heuristically assign a spin diffusion coefficient $D= \fr{\expec{r_n}^2}{30T_{2n}}$ where the $T_{2n}$ are evaluated here by only taking the dipolar contribution to the linewidth, $T_{2n}\app 1/\expec{d_{\CC}}$. Given a total time bounded by $T_1$, we can calculate the rms overall diffusion length~\cite{Zhang98} as $\xs=\sq{2DT_1}$ that is displayed as the blue points in \zfr{carbon}D. Also for reference is plotted the mean NV-NV distance $\app$12nm at 1ppm concentration, indicating that to a good approximation that the optically pumped polarization reaches to all parts of the diamond lattice between the NV centers.

\subsection{Lattice estimates for hyperfine couplings to NV and P1 reservoirs}
\zsl{NV}

Let us finally evaluate, through similar numerical means, details of the hyperfine interaction between $\Cs$ reservoir and the electron reservoirs of the P1 centers and NV centers. We draw a distinction between the NV and P1 centers in the fact that the former are spin-1, with a nonmagnetic $m_s=0$ state (with no hyperfine coupling to first order), while the latter are spin $1/2$. When hyperfine shifts exceed the observed 7T NMR linewidth $\xD f_{\R{det}}\sim$2kHz, it is safe to assume that these spins are unobservable - a case that is operational more strongly for the spin $1/2$ P1 centers.

In order to perform the estimation, in the generated lattice of size $\ell=\expec{r_e}$, we populate $\Cs$ spins with enrichment $\eta$, and include an electron at the lattice origin. The mean perpendicular hyperfine interaction between P1-$\Cs$ spins is calculated from the second moment, from the individual hyperfine couplings $A_{zx,j}$ that are smaller than the detection barrier $\xD f_{\R{det}}$
\bea
&&\expec{A_{zx}^{\obs}}= \lsb \sum_{j\in\obs}\expec{A^2_{zx,j}}\rsb^{1/2}\non\\
&=&\lsb\fr{1}{N_{\obs}}\sum_{j\in\obs}\lb\fr{\mu_0}{4\pi}\xg_e\xg_n\hbar\rb^2\fr{(3\sin\xt_j\cos\xt_j)^2}{r_j^6}\rsb^{1/2}\non
\eea
where  $N_{\obs}$ refers to the number of spins amongst the total $N=N_C\ell^3$ spins for which $\expec{A^2_{zx,j}}<(\xD f_{\R{det}})^2$. Here $r_j$ is the distance of the j$^{\R{th}}$ $\Cs$ nucleus, and $\xt_j$ the angle of P1-$\Cs$ axis to the magnetic field, and we have ignored the effect of $\Ns$ hyperfine interactions intrinsic to the P1 center. This effective hyperfine field, scaling with lattice enrichment $\eta$, is then indicated by the red (blue) points in \zfr{carbon}F for electron concentrations of 17ppm (48ppm) respectively. The error bars indicating the standard deviation of the obtained distributions upon several hundred realizations of the lattice. We observe that the effective hyperfine interaction $\expec{A_{zx}^{\obs}}$ is almost independent of $\eta$, and is higher for lattices with higher $P_e$ electron concentration. This is consistent with the results obtained through \zr{hyp_eqn} and matches our experimental observations in \zfr{enriched} of the main paper. For natural abundance samples we numerically obtain  $\expec{A_{zx}^{\obs}}=$1.4kHz,1.55kHz and 1.04kHz respectively for 17ppm, 48ppm, and 1ppm (representative of NV center concentrations), in agreement with estimates from \zr{hyp_eqn}.

Finally, let us estimate the number of spins that are directly polarized by the NV centers. In \zfr{carbon}G we evaluate the \I{full} hyperfine interaction to $\Cs$ spins of varying enrichment, considering no operational detection barrier.  \bea
&&\expec{A_{\NV}}= \lsb \sum_{j}\expec{A^2_{j,\NV}}\rsb^{1/2}\non\\
&=&\lsb\fr{1}{N}\sum_{j}\lb\fr{\mu_0}{4\pi}\xg_e\xg_n\hbar\rb^2\fr{\lsb(3r_{jz}^2 - 1)^2 + (3r_{jx}r_{jz})^2 + (3r_{jy}r_{jz})^2\rsb}{r_j^6}\rsb^{1/2}\non
\eea
where we employed a lattice size $\ell=\expec{r_{\NV}}$=12nm, and $N=N_C\ell^3$ refers to the number of $\Cs$ spins in the lattice with index $j$ running over all them. Here the angle part of the hyperfine interaction is evaluated by assigning the direction cosines, for instance as, $r_{jz}=(\vec{r_j}\cdot \hat{z}_{\NV})/r_j$, where $\hat{z}_{\NV}$ is the unit vector aligned along the N-V axis, collinear with the direction of the strong zero field splitting that forms the dominant part of the Hamiltonian at low fields. This effective hyperfine field, scaling with lattice enrichment $\eta$, is then indicated by the blue points in \zfr{carbon}G. Our DNP mechanism is a low-field one and is primarily effective when the full hyperfine coupling $\expec{A_{j,\NV}}$ is of the order of greater than the nuclear Larmor frequency $\xo_L=\xg_n\Bp$, where $\Bp$ is the polarizing field. We can heuristically measure the number of directly polarized spins surrounding an NV center as being those for which $\expec{A_{j,\NV}}>$200kHz. As \zfr{carbon}H indicates, the number of such directly polarized nuclei scales approximately linearly with $\Cs$ enrichment, with a constant ratio $\app 4.3\eta$ in the diamond lattice. Spin diffusion therefore plays an important role in the spread of polarization away from these directly polarized nuclei. %Ultimately at ultralow fields, this spin diffusion away in the inhomogenously polarized lattice masquerades as a T1 process.

\end{document}